\pdfoutput=1
\documentclass[11pt,a4paper]{article}

\usepackage{jheppub}
\usepackage[displaymath]{lineno}
\usepackage[T1]{fontenc} 

\usepackage{lineno}
\usepackage{atlasphysics}
\usepackage{graphicx}
\usepackage{amssymb}
\usepackage{array}
\usepackage{xspace}
\usepackage{enumerate}
\usepackage{multirow}
\usepackage{booktabs}
\usepackage{subfigure}
\usepackage{placeins}
\PassOptionsToPackage{hyphens}{url}\usepackage{hyperref}
\newcommand{\sqR}   {\tilde{q}^{}_{\mathrm{R}}}
\newcommand{\sqL}   {\tilde{q}^{}_{\mathrm{L}}}

\newcommand{\bone}   {\tilde{b}^{}_{1}}
\newcommand{\tone}   {\tilde{t}^{}_{1}}
\newcommand{\neut}  {\tilde{\chi}^{0}_{1}}  
\newcommand{\neuttwo}  {\tilde{\chi}^{0}_{2}}  
\newcommand{\chipm}  {\tilde{\chi}^{\pm}_{1}} 
\def\etmiss{\ensuremath{E_{\mathrm{T}}^{\mathrm{miss}}}\xspace}

\def\dphimin{\ensuremath{\mathrm{\Delta}\phi_{\mathrm{min}}^{\mathrm{4j}}}}
\def\meff{\ensuremath{\mathrm{m}_{\mathrm{eff}}}\xspace}

\def\bjets{$b$-jets\xspace}

\def\meff{\ensuremath{m_{\mathrm{eff}}}\xspace}
\def\meffe{\ensuremath{m_{\mathrm{eff}}^{\mathrm{4j}}}\xspace}
\def\meffi{\ensuremath{m_{\mathrm{eff}}^{\mathrm{incl}}}\xspace}
\def\mt{\ensuremath{m_{\mathrm{T}}}\xspace}
\def\metsige{\met/$\sqrt{H_{\rm T}^{\rm 4j}}$\xspace}

\newcommand{\gl}   {\tilde{g}}
\newcommand{\sq}   {\tilde{q}}

\renewcommand{\ttbar} {\ensuremath{t\bar{t}}\xspace}
\newcommand{\ttbb}{\ensuremath{t\bar{t} + b/b\bar{b}}\xspace}
\renewcommand{\met} {\ensuremath{E_{\mathrm{T}}^{\mathrm{miss}}}\xspace}
\renewcommand{\pt} {\ensuremath{p_\mathrm{T}}\xspace}

\newcommand{\lumi}{20.1}
\newcommand\totallumi{20.1~${\rm fb}^{-1}$}

\def\met{\ensuremath{E_{\mathrm{T}}^{\mathrm{miss}}}}

\newcommand{\PaperTitle}{Search for strong production of supersymmetric particles in final states with missing 
transverse momentum and at least three $b$-jets at $\sqrt{s}$~=~8~\TeV\ proton--proton collisions with the ATLAS detector}

\newcommand{\AbstractText}{This paper reports the results of a search for strong production of supersymmetric 
particles in 20.1~fb$^{-1}$~of proton--proton collisions  at a centre-of-mass energy of 8~\TeV\ using the ATLAS detector at the LHC. The search 
is performed separately in events with either zero or at least one high-$p_\mathrm{T}$ lepton (electron or muon), large missing transverse momentum, 
high jet multiplicity and at least three jets identified  as originated from the fragmentation of a $b$-quark. 
No excess is observed  with respect to the Standard Model predictions. The results 
are interpreted in the context of several supersymmetric models involving gluinos and scalar 
top and bottom quarks, as well as a mSUGRA/CMSSM model. 
Gluino masses up to 1340~\GeV\ are excluded, depending on the model, significantly 
extending the previous ATLAS limits.}

\usepackage{preprintcover}  
\PreprintCoverPaperTitle{\PaperTitle}  
\PreprintIdNumber{CERN-PH-EP-2014-110}  
\PreprintCoverAbstract{\AbstractText}  
\PreprintJournalName{JHEP}  

\begin{document}

\title{\boldmath \PaperTitle}

\author{The ATLAS collaboration}

\date{\today}

\begin{abstract}
  \AbstractText
\end{abstract}

\maketitle

\newpage
\section{Introduction}\label{sec-introduction}

Supersymmetry (SUSY)~\cite{Miyazawa:1966,Ramond:1971gb,Golfand:1971iw,Neveu:1971rx,Neveu:1971iv,Gervais:1971ji,Volkov:1973ix,Wess:1973kz,Wess:1974tw}  
 provides an extension of the Standard Model (SM) which can solve the hierarchy 
problem by 
introducing supersymmetric partners for the SM bosons and fermions~\cite{Dimopoulos:1981zb,Witten:1981nf,Dine:1981za,Dimopoulos:1981au,Sakai:1981gr,Kaul:1981hi}. 
In the framework of the 
$R$-parity-conserving minimal supersymmetric extension of the SM 
(MSSM)~\cite{Fayet:1976et,Fayet:1977yc,Farrar:1978xj,Fayet:1979sa,Dimopoulos:1981zb}, 
SUSY particles are produced in pairs and the lightest supersymmetric particle (LSP) 
is stable.  
In a large fraction of the MSSM R-parity conserving models, the LSP is the lightest neutralino ($\neut$)\footnote{The SUSY 
partners of the electroweak gauge 
and Higgs bosons are called gauginos and higgsinos, respectively. The charged gauginos
and higgsinos mix to form charginos ($\tilde{\chi}^\pm_i$, $i=1,2$), and the neutral ones mix to form 
neutralinos ($\tilde{\chi}^0_j$, $j=1,2,3,4$).} which is weakly interacting, thus providing a possible candidate for dark matter.
The coloured superpartners of quarks and gluons, the squarks ($\sq$) and gluinos ($\gl$), 
if not too heavy, would be produced in strong interaction processes at the Large Hadron 
Collider (LHC) and decay via cascades ending with the LSP. The undetected LSP results in 
missing transverse momentum -- whose magnitude is referred to as \met\ -- while the rest 
of the cascade yields final states with  multiple jets and possibly leptons.
The scalar partners of the right-handed and left-handed quarks, $\sqR$ and 
$\sqL$, mix to form two mass eigenstates $\tilde{q}_1$ and $\tilde{q}_2$. A substantial mixing is expected between 
$\tilde{t}^{}_{\mathrm{R}}$ and  $\tilde{t}^{}_{\mathrm{L}}$ because of the large Yukawa coupling of the top quark, leading to a large mass splitting between  
$\tilde{t}_1$ and $\tilde{t}_2$. 

SUSY can solve the hierarchy problem by preventing ``unnatural'' fine-tuning 
in the Higgs sector provided that the superpartners of the top quark  have 
masses not too far above the weak scale~\cite{Barbieri:1987fn,deCarlos:1993yy}. This condition 
requires that the gluino is not too heavy   in order to limit its contribution  to 
the radiative corrections to the stop masses. Besides, the mass of the left-handed sbottom  ($\tilde{b}_\mathrm{L}$) is tied to the stop mass because of the SM weak isospin symmetry.  
As a consequence, the lightest sbottom ($\bone$) and stop ($\tone$) could be produced via strong production with relatively large 
cross-sections at the LHC, mainly via direct pair production or through $\gl\gl$ production followed 
by $\gl \rightarrow \bone b$ or $\gl \to \tone t$ decays.

This paper presents new results of a search for supersymmetry in final states with large \met\ and at least three 
jets identified as originated from the fragmentation of a $b$-quark ($b$-jets). 
The previous version  of this analysis, using only events with no electrons or muons (0-lepton) in the final state, was 
performed with the full data set recorded by the ATLAS detector in 2011 
at a centre-of-mass energy of 7~\TeV~\cite{3bjets_EPJC}.  
The present analysis uses the dataset of \lumi~$\ifb$ collected during 
2012 at a centre-of-mass energy of 8 \TeV, and extends the previous search 
by considering events with at least one high-\pt\ electron or muon (1-lepton) in the final state.  

The results are interpreted in the context of various SUSY models where top or bottom quarks are produced in 
gluino decay chains. Additional interpretations are provided for a direct sbottom pair production scenario 
where the sbottom decays into a bottom quark and the next-to-lightest neutralino, $\neuttwo$, followed by 
the $\neuttwo$ decay into a Higgs boson and the LSP, and for a mSUGRA/CMSSM model designed to accommodate a 
Higgs boson with a mass of about 125~\GeV. Exclusion limits in similar SUSY models have been placed by 
other analyses carried out by the ATLAS~\cite{Aad:2013wta,Aad:2014pda} and CMS~\cite{Chatrchyan:2013wxa,Chatrchyan:2013iqa,Chatrchyan:2013fea,Chatrchyan:2014lfa} 
collaborations with the same integrated luminosity at a centre-of-mass energy of 8~\tev.

\section{SUSY signals}\label{sec-susy}

In order to confront the experimental measurements with theoretical expectations, several classes of simplified models with $b$-quarks in the final state are considered. 
Results from the 0-lepton channel are used to explore all models considered, while the complementarity between the searches in the 0- and 
1-lepton channels is used to maximise the sensitivity to models predicting top quarks in the decay chain.

In the first class of simplified models, the lightest stops and sbottoms are lighter 
than the gluino, such that $\bone$  and $\tone$ are produced either in pairs, or via 
gluino pair production followed by  $\gl \rightarrow \bone b$ or $\gl \rightarrow \tone t$ 
decays. The mass of the $\chipm$ is set at 60~\GeV\ consistently for all models. The following models, 
 also shown in figure~\ref{f-feyn1}, are considered: 

\begin{figure*}[hbpt]
\centering
\subfigure[]{\includegraphics[width=0.37\columnwidth]{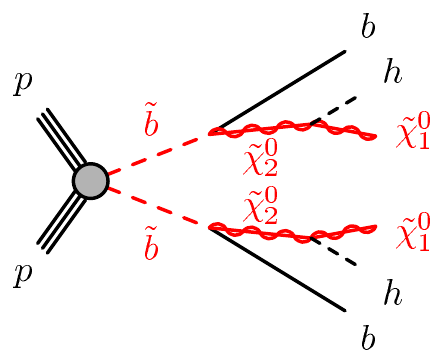}}\\
\subfigure[]{\includegraphics[width=0.37\columnwidth]{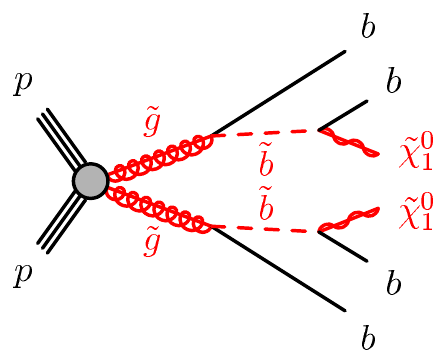}}
\subfigure[]{\includegraphics[width=0.37\columnwidth]{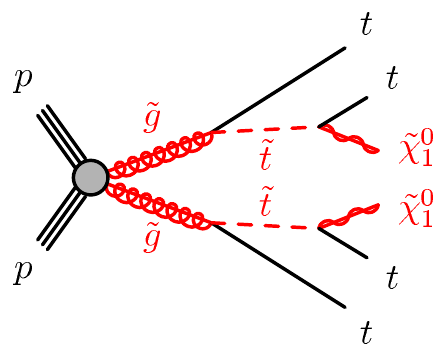}}
\caption{This figure shows the diagrams for the (a) direct--sbottom, (b) gluino--sbottom 
and (c) gluino--stop  scenarios studied in this paper. 
The different decay modes are discussed in the text.}\label{f-feyn1}
\end{figure*}

\begin{itemize}
\item {\bf Direct--sbottom model:} in this model, the $\bone$ is produced in pairs and is assumed to decay exclusively via 
$\bone \rightarrow b+\neuttwo$. The slepton masses are set above a few \TeV\ and only  
the configuration $m_{\neuttwo} > m_{\neut} + m_h$ with a branching 
ratio for $\neuttwo \rightarrow h + \neut$ of 100\% is considered. The mass of the lightest 
neutral Higgs boson $h$ is set to 125~\GeV, 
and its decay branching ratios are assumed to be those of the SM Higgs boson.  
The analysis is mainly sensitive to signal events where both Higgs bosons decay into a $b\bar{b}$ pair, yielding six $b$-quarks, two neutralinos and 
no leptons at the end of the decay chain.

\item {\bf Gluino--sbottom model:}  in this model, the $\bone$ is the lightest squark, all other squarks are heavier than the gluino, and 
$m^{}_{\gl}>m^{}_{\bone}+m_b$ such that the branching ratio for $\gl \rightarrow \bone b$ decays is 100\%. 
Sbottoms  are produced 
 in pairs or via gluino pair production and are assumed to decay exclusively via $\bone \rightarrow b\neut$. The analysis is  
sensitive to the gluino-mediated production, which has  four bottom quarks, two neutralinos and no leptons at the end of the decay chain.

\item {\bf Gluino--stop models:}  in these models, the $\tone$ is the lightest squark, all other squarks are heavier than the gluino, and 
$m^{}_{\gl}>m^{}_{\tone}+m_t$ such that the branching ratio for $\gl \rightarrow \tone t$ decays is 100\%. 
Stops are produced in pairs or via gluino pair production and are assumed to decay 
exclusively via $\tone \rightarrow b \chipm$ (model I), or via $\tone \rightarrow t \neut$ (model II). For the first model, 
the chargino mass is assumed to be twice the mass of the neutralino, such that the chargino decays into a neutralino and a virtual $W$ boson. 
The analysis is sensitive to the gluino-mediated production with 
two top quarks, two bottom quarks, two virtual $W$ bosons and two neutralinos (model I), or four top quarks and two neutralinos (model II) at the end of the SUSY decay chain, yielding signatures with or without leptons. 

\end{itemize}

In the second  class of simplified models, all sparticles, apart from the 
gluino and the neutralino, have masses well above the \TeV\ scale such that 
the $\tone$ and the $\bone$ are only produced off-shell via prompt decay of the gluinos.
Thus, the sbottom and stop masses have little impact on the kinematics of the final state. 
The following models,  also shown in figure~\ref{f-feyn2}, are considered: 

\begin{figure*}[hbpt]
\centering
\subfigure[]{\includegraphics[width=0.37\columnwidth]{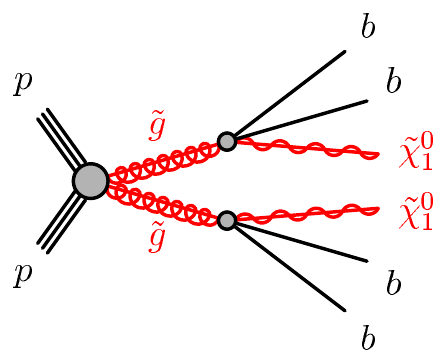}}
\subfigure[]{\includegraphics[width=0.37\columnwidth]{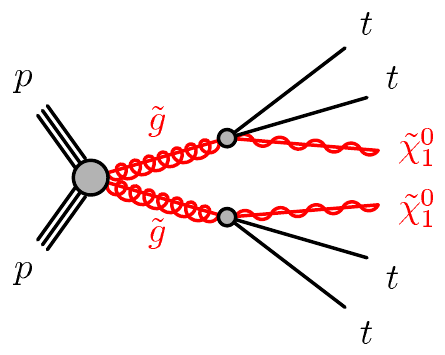}}\\
\subfigure[]{\includegraphics[width=0.37\columnwidth]{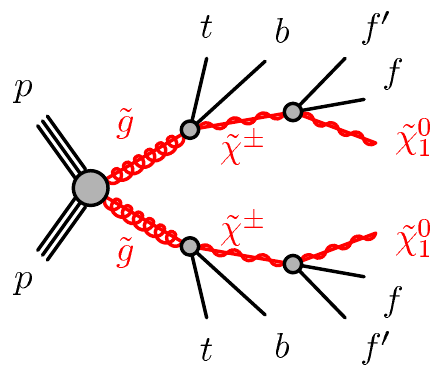}}
\caption{This figure shows the diagrams for the (a) Gbb, (b) Gtt  
and (c) Gtb  scenarios studied in this paper. 
The different decay modes are discussed in the text.}\label{f-feyn2}
\end{figure*}

\begin{itemize} 

\item {\bf Gluino--sbottom off-shell (Gbb) model:}  in this model, the $\bone$ is the lightest squark, but with $m^{}_{\gl}<m^{}_{\bone}$. 
A three-body decay $\gl \rightarrow b\bar{b}\neut$ via an off-shell sbottom is assumed for the gluino with a branching 
ratio of 100\%.  As for the gluino--sbottom model,  four bottom quarks, two neutralinos and no leptons are expected at the end of the decay chain. Therefore, only the 0-lepton analysis is
used for the interpretation. 

\item {\bf Gluino--stop off-shell (Gtt) model:}  in this model, the $\tone$ is the lightest squark, but $m^{}_{\gl}<m^{}_{\tone}$. 
A three-body decay $\gl \rightarrow t\bar{t}\neut$ via an off-shell stop is assumed for the gluino with a branching 
ratio of 100\%.  Four top quarks and two neutralinos are expected as decay products of the two gluinos,  
resulting in signatures with or without leptons. 

\item {\bf Gluino--stop/sbottom off-shell (Gtb) model:}  in this model, the $\bone$ and $\tone$ are the lightest squarks, with 
$m^{}_{\gl}<m^{}_{\bone, \tone}$. Pair production of gluinos is the only process taken 
into account, with gluinos decaying via virtual stops or sbottoms, with a branching ratio 
of 100\% assumed for both $\tone \rightarrow b+\chipm$ and $\bone \rightarrow t+\chipm$. 
The mass difference between charginos and neutralinos is set to 2~\GeV, 
such that the fermions produced in $\chipm \rightarrow \neut + ff'$ do not contribute  to 
the event selection, and gluino decays result in effectively three-body decays  
($bt\neut$).  Two top quarks,  two bottom quarks and two neutralinos are expected as decay products of the two gluinos, yielding signatures with or without leptons.

\end{itemize}

The results are also interpreted in the context of a minimal supergravity model \\
{\bf mSUGRA/CMSSM}~\cite{Chamseddine:1982,Barbieri:1982eh,Ibanez:1982ee,Hall:1983iz,Ohta:1982wn, Kane:1993td} 
specified by five parameters: the universal scalar mass $m_0$, 
the universal gaugino mass $m_{1/2}$, the universal trilinear scalar coupling $A_0$, 
the ratio of the vacuum expectation values of the two Higgs fields $\tan \beta$, 
and the sign of the higgsino mass parameter $\mu$. The model used for interpretation has 
$A_0 = -2 m_0$, $\tan \beta = 30$, $\mu > 0$ and is designed to accommodate a SM Higgs boson with a mass of around 125~\GeV,
in the $m_0 - m_{1/2}$ range relevant for this analysis.

\section{The ATLAS detector and data sample}\label{sec-detector}

The ATLAS detector~\cite{DetectorPaper:2008} is a multi-purpose particle physics
detector with forward-backward symmetric cylindrical geometry.\footnote{ATLAS uses a right-handed coordinate 
system with its origin at the nominal interaction point (IP) in the centre of the detector and the $z$-axis along 
the beam pipe. The $x$-axis points from the IP to the centre of the LHC ring, and the $y$-axis points upward. 
Cylindrical coordinates $(r,\phi)$ are used in the transverse plane, $\phi$ being the azimuthal angle around the 
beam pipe. The pseudorapidity is defined in terms of the polar angle $\theta$ as $\eta=-\ln\tan(\theta/2)$, and the 
distance~$\mathrm{\Delta} R$ in the ($\eta$,$\phi$) space is defined as 
$\mathrm{\Delta} R = \sqrt{(\mathrm{\Delta}\eta)^2+(\mathrm{\Delta}\phi)^2}$.}    
It consists of inner tracking devices
surrounded by a superconducting solenoid, electromagnetic
and hadronic calorimeters and a muon spectrometer with a magnetic field produced by three large 
superconducting toroids each with eight coils.  The inner detector,
in combination with the 2~T field from the solenoid,
provides precision tracking of charged particles for $|\eta| < 2.5$.
 It consists of a silicon pixel detector, a silicon microstrip
detector and a straw-tube tracker that also provides transition
radiation measurements for electron identification.
The calorimeter system covers the pseudorapidity range
$|\eta| < 4.9$. A high-granularity liquid-argon (LAr) sampling calorimeter with lead absorber  
is used to measure the energy of electromagnetic (EM) showers 
within $|\eta| < 3.2$. Hadronic showers are measured by an iron/scintillator tile calorimeter in the central region ($|\eta| < 1.7$) and 
by a LAr calorimeter in the end-cap ($1.5<|\eta|<3.2$).  
The forward region ($3.1<|\eta|<4.9$) is instrumented with a LAr calorimeter for both EM and hadronic measurements. 
The muon spectrometer has separate trigger and high-precision 
tracking chambers, which provide muon identification and momentum measurement 
for $|\eta| < 2.7$.                                
                                                                                 
The data sample used in this analysis was recorded during the period from March 2012
to December 2012 with the LHC operating at a $pp$ centre-of-mass energy of 8~\TeV.
After the application of the data-quality requirements, 
the total integrated luminosity amounts to~\totallumi, with an associated uncertainty of $\pm$2.8\% measured
using techniques similar to those detailed in ref.~\cite{Aad:2013ucp}, resulting from a preliminary calibration of the luminosity scale using beam-separation scans performed in November 2012.  
Events for the analysis are selected using a trigger based on a missing 
transverse momentum selection, which is found to be $> 99$\% efficient after the
offline requirements $\met >150$~\gev\ and at least one reconstructed jet of transverse
momentum \pt$>90$~\gev\ and $|\eta| < 2.8$.

\section{Simulated event samples}\label{sec-samples}

Samples of simulated Monte Carlo (MC) events are used to assess the sensitivity 
to specific SUSY models and aid in the prediction of the SM backgrounds. 
Jets are labelled as true $b$-jets in MC simulations  
if they satisfy the kinematic requirements applied to $b$-jets detailed in section~\ref{sec-objrec} and if they 
are matched to a generator-level $b$-quark with \pt\ $>$ 5 \gev\ within $\mathrm{\Delta} R =0.3$. 
 The various background processes are then classified into two categories:  
 those leading to final states with at least three true $b$-jets form the irreducible component  
 while all other processes form the reducible component, the latter being the dominant source of background.   
Irreducible backgrounds arise mainly from $\ttbar + b$ and $\ttbar + b\bar{b}$ production, and to a minor extent from $\ttbar$+$Z/h$ followed by 
$Z/h \rightarrow b\bar{b}$. Their contributions are estimated from MC simulations that are generated inclusively, 
each event being classified at a later stage based on the number of true \bjets found. 
Contributions from background events in which at least one jet is misidentified as a $b$-jet  arise mainly from
\ttbar production in association with light-parton- and $c$-jets. Sub-dominant contributions arise from \ttbar\ production in association with $W/Z/h$+jets (except events with $Z/h \rightarrow b\bar{b}$), single top quark production,
$W/Z$+jets production, and diboson ($WW,~WZ,~ZZ$) production. 
The contributions from all these reducible background processes are estimated simultaneously using a data-driven method described in section~\ref{sec-MM}, and  
MC samples are only used for comparison. Details of the MC simulation samples used in this analysis, as well as the order of 
cross-section calculations in perturbative QCD (pQCD) used for yield normalisation, are shown in table~\ref{t-MCsamples}. The background prediction calculated as the sum of the event yield predicted by the 
MC simulation for each SM process is referred as the MC-only prediction in the following.

\begin{table}[h]

\centering
{\tiny
\begin{tabular}{ccccc}
\toprule
Process & Generator & Cross-section  & Tune   & PDF set  \\
 & + fragmentation/hadronisation       & order  &  & \\
\midrule
{\bf \ttbar} & {\textsc POWHEG-r2129}~\cite{Nason:2004rx,Frixione:2007vw,Alioli:2010xd} & NNLO+NNLL~\cite{Cacciari:2011hy,Baernreuther:2012ws,Czakon:2012zr,Czakon:2012pz,Czakon:2013goa,Czakon:2011xx} & {\textsc PERUGIA2011C}~\cite{Skands:2010ak} & {\textsc CT10}~\cite{Lai:2010vv} \\%
 & + {\textsc PYTHIA-6.426}~\cite{pythia} & & & \\%

{\bf \ttbar}* & {\textsc POWHEG-r2129} & NNLO+NNLL & {\textsc AUET2B}~\cite{tuneAUET2B} & {\textsc CT10} \\
& + {\textsc HERWIG-6.520}~\cite{Corcella:2000bw} & &  &  \\

 {\bf \ttbar}* & {\textsc MadGraph-5.1.5.11}~\cite{madgraph} & NNLO+NNLL & {\textsc PERUGIA2011C}& {\textsc CT10} \\
  &  + {\textsc PYTHIA-6.427} & & & \\
  
\midrule
{\bf Single top} & & & & \\
        $t$-channel & {\textsc AcerMC-3.8}~\cite{Kersevan:2004yg} & NNLO+NNLL~\cite{Kidonakis:2011wy}  & {\textsc AUET2B} & {\textsc CTEQ6L1}~\cite{Pumplin:2002vw} \\
        & + {\textsc PYTHIA-6.426} &  &  & \\
                
        $s$-channel, $Wt$ &  {\textsc MC@NLO-4.06}~\cite{Frixione:2002ik,Frixione:2005vw} & NNLO+NNLL~\cite{Kidonakis:2010tc, Kidonakis:2010ux} & {\textsc AUET2B} & {\textsc CT10 } \\
        & + {\textsc HERWIG-6.520} &  & & \\        
\midrule
{\bf Top+Boson} & & & & \\
         $\ttbar+W$, $\ttbar+Z$ & {\textsc MadGraph-5.1.4.8}  &  NLO~\cite{Garzelli:2012bn} & {\textsc AUET2B} & {\textsc CTEQ6L1} \\
           & + {\textsc PYTHIA-6.426} &  &  & \\
         
         $\ttbar+h$  &  {\textsc MadGraph-5.1.4.8}   & NNLO ~\cite{Dittmaier:2012vm} & {\textsc AU2}~\cite{tuneAU2} & {\textsc CTEQ6L1} \\
             &  +  {\textsc PYTHIA-8.165}~\cite{pythia8} & & &  \\
           
\midrule
{\bf \boldmath $W$+jets, $Z$+jets} & {\textsc SHERPA-1.4.1} & NNLO~\cite{Catani:2009sm} &{\textsc AUET2B} & {\textsc CT10}  \\
 & & with {\textsc MSTW2008 NNL0}~\cite{Martin:2009iq} & & \\
\midrule
{\bf Dibosons} & & & & \\
     $WW$, $WZ$, $ZZ$ & {\textsc SHERPA-1.4.1} & NLO~\cite{Binoth:2006mf} & {\textsc AUET2B} & {\textsc CT10} \\
\bottomrule
\end{tabular} 
}       
\caption{List of MC generators used for the different background processes. Information is given about the pQCD highest-order accuracy used for the normalisation of the different samples, the underlying event tunes and PDF sets considered. Samples labelled with * are employed for the evaluation of systematic uncertainties.}
  \label{t-MCsamples} 
\end{table}

The SUSY signal samples used in this analysis were generated with {\textsc Herwig++ 2.5.2}~\cite{Bahr:2008pv}. 
For the Gbb model, in order to ensure an accurate treatment of the initial-state 
radiation (ISR), {\textsc MadGraph-5.1.5.4} interfaced to {\textsc PYTHIA-6.426} is used. 
All the signal samples were generated with the parton distribution function (PDF) set {\textsc CTEQ6L1}. 
They are normalised to the signal cross-sections calculated to next-to-leading order in the strong coupling constant, adding the re-summation of 
soft gluon emission at next-to-leading-logarithmic approximation  
(NLO+NLL)~\cite{Beenakker:1996ch,Kulesza:2008jb,Kulesza:2009kq,Beenakker:2009ha,Beenakker:2011fu}. 

The nominal cross-section and the uncertainty $\sigma^{\mathrm{SUSY}}_{\mathrm{theory}}$ are taken 
from an envelope of cross-section predictions using different PDF sets and factorisation and 
renormalisation scales, as prescribed in ref.~\cite{Kramer:2012bx}. 
An additional source of systematic uncertainty is taken into account for the Gbb model, where the modelling of the ISR can significantly affect the signal acceptance in the region of the 
parameter space with  small mass splitting $\Delta m$ between the $\gluino$ and the $\neut$.   
The uncertainty on the signal acceptance is estimated by varying the value of $\alpha_{\mathrm{S}}$, the renormalisation and factorisation scales, as well as the  matching parameters in the 
{\textsc MadGraph+PYTHIA-6} MC samples. This uncertainty amounts to 30\% for the lowest mass splitting and decreases exponentially with increasing $\Delta m$. It is negligible in the region with $\Delta m > 200$~\gev, where the predictions from  {\textsc MadGraph+PYTHIA-6}  and {\textsc Herwig++} are consistent within statistical uncertainties. 
This systematic uncertainty is negligible for all other signal models considered in this paper. 

All the MC samples are processed either through a full 
simulation of the ATLAS detector~\cite{atlassimulation} based on {\textsc GEANT4}~\cite{geant4} 
or a fast simulation~\cite{atlfast} that uses a parameterisation of the performance of the ATLAS electromagnetic and hadronic calorimeters 
and GEANT4 elsewhere.   
Potential differences between the full and fast simulations were found negligible for this analysis.  
The effect of multiple $pp$ interactions in the same or neighbouring bunch crossings  (pile-up) is 
incorporated into the simulation by overlaying additional minimum-bias 
events generated with {\textsc PYTHIA-8} onto the hard-scattering process.  Simulated events are then weighted to match the observed distribution 
of the number of $pp$ interactions, and are reconstructed in exactly the same way as the data otherwise. 

\section{Object reconstruction and identification}\label{sec-objrec}

Jets are reconstructed from three-dimensional calorimeter energy clusters with 
the \mbox{anti-$k_t$} jet algorithm~\cite{Cacciari:2008gp} with a radius 
parameter $R=0.4$. The measured jet energy is corrected for inhomogeneities and 
for the non-compensating response of the calorimeter by differently weighting 
energy deposits arising from electromagnetic and hadronic showers with correction 
factors derived from MC simulations and {\it in situ} measurement in data~\cite{JES}.  
Jets are corrected for pile-up using a method proposed in 
ref.~\cite{Cacciari:2007fd}. Finally, additional corrections are applied to calibrate the
jet energy to  the energy of the corresponding jet of stable particles.
Only jets with $|\eta| < 4.5$ and $\pT > 20$~\GeV\ after 
calibration are retained. 

To remove events with jets from detector noise and non-collision backgrounds,  events are rejected  
if they include jets failing  to satisfy  the loose quality criteria described in ref.~\cite{JES}. 
Additional cleaning cuts based on the fraction of the transverse momentum of the jet carried by reconstructed charged particle tracks and the fraction of the jet energy in the EM calorimeter are applied to reject events containing spurious jet signals. Except for the \met\ computation, only jets with $|\eta| < 2.8$ are further considered.

A neural-network-based algorithm~\cite{btag1} is used to identify jets originated from the fragmentation of a $b$-quark. 
It uses as inputs the output weights of different algorithms 
exploiting the impact parameter of the inner detector tracks, the secondary vertex 
reconstruction and the topology of $b$- and $c$-hadron decays inside the jet. 
The algorithm used has an efficiency of 70\% for tagging \bjets in a MC sample of 
\ttbar\ events with rejection factors of 137, 5 and 13 against light-quarks, 
$c$-quarks and $\tau$ leptons respectively. The $b$-jets are identified within the 
 acceptance of the inner detector ($|\eta| < 2.5$). To compensate for the small 
differences between the $b$-tagging efficiencies and the misidentification (mistag) 
rates in data and MC simulations, correction factors are applied to each jet in the simulations, 
as described in refs.~\cite{btag1,btag2,btag3,btag4}. These corrections are of the order of a few per cent.

Electrons are reconstructed from energy clusters in the electromagnetic calorimeter 
associated with tracks in the inner detector. Electron
candidates are required to have \mbox{$\pt > 20$~\GeV} and $|\eta|<2.47$, and
must satisfy the {\it medium} shower shape and track selection criteria based upon those described in
ref.~\cite{Aad:2014fxa}, adapted for 2012 data conditions. Muon candidates are identified using a match between an extrapolated inner 
detector track and one or more track segments in the muon spectrometer~\cite{ref-mu}, 
and are required to have $\pt>10$~\GeV\ and $|\eta| < 2.5$. In order to reduce the 
contributions from semileptonic decays of hadrons, 
lepton candidates  found within $\mathrm{\Delta} R = 0.4$ of a jet are discarded. 
Events containing one or more muon candidates 
that have a transverse (longitudinal) impact parameter $d_0$ ($z_0$) with respect to the primary vertex 
larger than 0.2 (1)~mm  are rejected to suppress cosmic rays.
{\it Signal} electrons (muons) are required to be isolated, i.e. 
the  sum of the extra transverse energy deposits in the calorimeter, corrected for pile-up effects,   
within a cone of $\Delta R=0.3$ around the lepton candidate 
must be less than 18\% (23\%) of the lepton \pt, and 
the scalar sum of the transverse momenta of tracks within a cone of $\Delta R=0.3$  around the 
lepton candidate must be less than 
16\% (12\%) of the  lepton \pt. Energy deposits and tracks of the leptons themselves  are not included.
In addition, to further suppress leptons originating from secondary vertices, signal electrons 
(muons) must have $|z_0 \sin\theta|<0.4$~mm and $d_0 / \sigma_{d_0} < 5 (3)$. 
Signal electrons must also satisfy tighter quality requirements based upon the criteria 
denoted by {\it tight} in ref.~\cite{Aad:2014fxa}. 
Correction factors are applied to MC events to match the lepton identification and reconstruction 
efficiencies observed in data. 

The measurement of the missing transverse momentum vector 
(and its magnitude \etmiss) is based on the transverse momenta of all jets, 
electron and muon candidates, and all                                                                                        
calorimeter clusters not associated with such objects. Clusters associated with either
electrons or photons with $\pt >10 $~\GeV, and those associated with jets with
$\pt >20 $~\GeV, make use of the calibrations of these respective objects. 
Clusters not associated with these objects are calibrated using both
calorimeter and tracker information~\cite{Aad:2012re}.

\section{Event selection}\label{sec-sel}

Following the trigger and object selection requirements described in sections~\ref{sec-detector} and \ref{sec-objrec},
events are discarded if they fail to satisfy basic quality criteria designed to reject detector noise and 
non-collision backgrounds. Candidate events are required to have a reconstructed primary vertex 
associated with five or more tracks with $\pt >0.4$~\GeV\ \cite{Aad:2013zwa}; when more than one such vertex 
is found, the vertex with the largest summed $\pt^2$ of the associated tracks is chosen 
as the primary vertex. Events must have $\met>150$~\GeV\ and at least four jets with $\pt>30$~\GeV. 
The leading jet is required to have $\pt>90$~\GeV\ and at least three of the  jets with $\pt>30$~\GeV\ must be $b$-tagged.
The events selected at this stage are then divided into two complementary channels 
based on the number of leptons: i) 0-lepton channel, formed by events with no reconstructed electron or muon candidates; and
ii) 1-lepton channel, formed by events with at least one signal lepton with $\pt>25$~\GeV.
After this basic selection, events are classified into several signal regions (SR), designed 
to provide sensitivity to the different kinematic topologies associated with the various 
SUSY models under study. Each SR is defined by a set of selection criteria using additional 
event-level variables calculated from the reconstructed objects. 

For the 0-lepton channel, four additional variables are used:

\begin{itemize}

\item The inclusive effective mass \meffi, defined  as the scalar sum of the \met\ and the $\pt$ 
of all jets with $\pt >$ 30~\gev. It is correlated with the overall mass scale of the  hard-scatter interaction 
and provides good discrimination against SM background. 

\item The exclusive effective mass \meffe, defined as the scalar sum of the \met\ and the $\pt$ 
of the four leading jets. It is used to suppress the multi-jet background 
and to define the SRs targeting SUSY signals where exactly four $b$-jets and large \met\ are expected in the final state. 

\item \dphimin, defined as the minimum azimuthal separation between any of the four leading 
jets and the missing transverse momentum direction. To remove multi-jet events where \met\ results
from poorly reconstructed jets or from neutrinos emitted close to the direction of the jet axis, 
events are required to have $\dphimin>0.5$ and $\met/\meffe > 0.2$. The combination of these
two requirements reduces the contribution of the multi-jet background to a negligible amount. 

\item The missing transverse momentum significance, defined as \metsige,
where $H_{\rm T}^{\rm 4j}$  is the scalar sum of the transverse momenta of the four leading jets, 
is used to define the SRs aiming at SUSY signals with four jets in the final state.

\end{itemize}

For the 1-lepton channel, event selections are defined using the following variables:

\begin{itemize}
 
\item \meffi, defined as for the 0-lepton channel with the addition of the \pt\ of all  signal 
leptons with $\pt>20$ \GeV.

\item The transverse mass \mt\, computed from the leading lepton and the missing transverse
momentum as $\mt=\sqrt{2 p_{\mathrm{T}}^\ell \met(1-\cos\Delta\phi(\ell,\met))}$. It is used to reject the 
main background from \ttbar\ events where one of the $W$ bosons decays leptonically. 
After the \mt\ requirement, the dominant 
contribution to the \ttbar\ background in the 1-lepton channel  arises from dileptonic \ttbar\ events.

\end{itemize}

The baseline event selections for each channel and the nine resulting SRs are summarised in table~\ref{t-SR}.
Three sets of SRs, two for the 0-lepton channel and one for the 1-lepton channel, each
denoted by `0$\ell$' or `1$\ell$', respectively,
 are defined to enhance the sensitivity to the various models considered. 
They are characterised by having relatively hard \met\ requirements and at least four (SR-0$\ell$-4j),
six (SR-1$\ell$-6j) or seven (SR-0$\ell$-7j) jets, amongst which at least three are $b$-jets. Signal
regions with zero leptons and at least four jets target SUSY models with sbottoms  
in the decay chain, while the 1-lepton and the 0-lepton--7-jets SRs aim to probe
SUSY models predicting top quarks in the decay chain. 
All SRs are further classified as A/B/C depending on the thresholds applied to 
the various kinematic variables previously defined, designed to target different mass hierarchies
in the various scenarios considered. In particular, a dedicated SR aiming to
increase the sensitivity at low mass splitting between the gluino and the $\neut$ in the Gbb model is defined. 
 This SR (denoted by SR-0$\ell$-4j-C* in table~\ref{t-SR}) exploits the recoil against an ISR jet by requiring 
the leading jet to fail the $b$-tagging requirements.

 \begin{table*}[h]
    \centering
       \begin{tabular}{lcccc}
        \toprule
\multicolumn{5}{c}{Baseline 0-lepton selection: lepton veto, $p_{\mathrm{T}}^{j_1} > 90$ \gev, \met $>$ 150 \gev, }\\
\multicolumn{5}{c}{$\geq$ 4 jets with $\pt >$ 30 \gev, $\dphimin>0.5$, $\met/\meffe>0.2$, $\geq$ 3 $b$-jets with $\pt >$ 30 \gev} \\
        \midrule 
	& $N$ jets (\pt\ [\gev]) &  \met\ [\gev] & \meff\ [\gev] & \metsige\ [$\sqrt{\gev}$]\\
        \midrule
        SR-0$\ell$-4j-A & $\geq$ 4 (50)  & $>$ 250  & \meffe $>$ 1300  & -- \\
        SR-0$\ell$-4j-B & $\geq$ 4 (50)  & $>$ 350  & \meffe $>$ 1100  & -- \\       
        SR-0$\ell$-4j-C* & $\geq$ 4 (30)  & $>$ 400  & \meffe $>$ 1000  & $>16$ \\             
         \midrule         
        SR-0$\ell$-7j-A & $\geq$ 7 (30)  & $>$ 200  & \meffi $>$ 1000  & -- \\       
        SR-0$\ell$-7j-B & $\geq$ 7 (30)  & $>$ 350  & \meffi $>$ 1000  & -- \\       
        SR-0$\ell$-7j-C & $\geq$ 7 (30)  & $>$ 250  & \meffi $>$ 1500  & -- \\              
        \bottomrule
	\multicolumn{5}{c}{Baseline 1-lepton selection: $\geqslant$ 1 signal lepton ($e$,$\mu$),  $p_{\mathrm{T}}^{j_1} > 90$ \gev, \met $>$ 150 \gev, } \\
\multicolumn{5}{c}{$\geq$ 4 jets with $\pt >$ 30 \gev, $\geq$ 3 $b$-jets with $\pt >$ 30 \gev } \\
        \toprule 
	& $N$ jets (\pt\ [\gev]) &  \met\ [\gev]   & $\mt$ [\gev]  &  \meffi [\gev]   \\
         \midrule
	SR-1$\ell$-6j-A & $\geq$ 6 (30) & $>$ 175  & $>$ 140 & $>$ 700  \\       
        SR-1$\ell$-6j-B & $\geq$ 6 (30) & $>$ 225  & $>$ 140   & $>$ 800 \\       
        SR-1$\ell$-6j-C & $\geq$ 6 (30) & $>$ 275  & $>$ 160  & $>$ 900  \\   
	\bottomrule   
             \end{tabular}
\caption{Definition of the signal regions used in the 0-lepton and 1-lepton selections.  The 
jet \pt\ threshold requirements are also applied to $b$-jets.  
The notation SR-0$\ell$-4j-C* means that the leading jet is required to fail the $b$-tagging requirements
 to target the region close to the kinematic boundary in the Gbb model.}
   \label{t-SR}
 \end{table*}

\section{Background estimation} \label{sec-bkg}

The main source of reducible background is the production of \ttbar\ events 
where a $c$-jet or a hadronically decaying $\tau$ lepton  is mistagged as a $b$-jet, the contribution from \ttbar\ events with a light-quark or 
gluon jet mistagged as a $b$-jet being relatively small.  In the 0-lepton channel, most of these 
\ttbar\ events have a $W$ boson decaying leptonically where the lepton is not reconstructed, 
is outside the acceptance, is misidentified as a jet, or is a $\tau$ lepton which decays hadronically. 
In the 1-lepton channel, the high \mt\ requirement used to define the SRs 
enhances the contribution from  dileptonic \ttbar\ events, where one of the two leptons is a hadronically decaying $\tau$ lepton.
 Additional minor sources of reducible background are single-top production, $\ttbar$+$W$/$Z$/$h$ (except events with $Z/h\rightarrow b\bar{b}$), 
$W$/$Z$+heavy-flavour jets, and diboson events. The irreducible backgrounds with at least three 
true $b$-jets in the final state arise predominantly from \ttbb\ events, and to
a minor extent from $\ttbar$+$Z$/$h$ production with a subsequent decay of the $Z$ or  
 Higgs boson into a pair of $b$-quarks. 

Different techniques are used to estimate the contribution from the reducible and the 
irreducible backgrounds in the SRs, explained in detail in the following sections.

\subsection{Reducible background} \label{sec-MM}

All reducible backgrounds are estimated simultaneously using a data-driven method 
which predicts the contribution from events with at least one mistagged jet amongst the three selected $b$-jets. 
This estimate is based on a matrix method (MM) similar to that used in ref.~\cite{Aad:2010ey} to predict the contribution from background events 
with fake and non-prompt leptons. It consists of solving a system of equations 
relating the number of events with $N_{\mathrm{j}}$ jets and $N_b$ $b$-jets to the number of events with $N_b^{\mathrm{T}}$ true 
$b$-jets and $(N_{\mathrm{j}} - N_b^{\mathrm{T}})$ non-true $b$-jets, prior to any $b$-tagging requirement. This method is applied on an event-by-event basis, 
such that for a given event containing $N_j$ jets satisfying the $\eta$ and $\pt$ requirements applied to $b$-jets, 
 $2^{N_j}$ linear equations are necessary to take into account the possibility for each of the $N_j$ jets to be a true $b$-jet or not. 
These linear equations are written in the form of a matrix of dimension $2^{N_{\mathrm{j}}} \times 2^{N_{\mathrm{j}}}$, the elements of which 
are functions of the probabilities for each jet in the event to be tagged or mistagged as a $b$-jet. 
The system of $2^{N_j}$ equations is solved by inverting the matrix, and an event weight is calculated from the combinations  
containing zero, one or two true $b$-jets. The weights obtained for each event satisfying all selection criteria except the $b$-tagging requirements are then 
summed to obtain the predicted number of events with at least one mistagged $b$-jet amongst the selected $b$-jets.

The $b$-tagging efficiencies used in the MM are measured in data for each jet-flavour using different techniques~\cite{btag1,btag2,btag3}. 
They are labelled as $\epsilon_b$, $\epsilon_c$, $\epsilon_\tau$ and  $\epsilon_{\mathrm{l}}$ for $b$-, $c$-, $\tau$- and light-parton-jets respectively. 
However, since the origin of a jet candidate is unknown in data, a mistag rate based on MC simulations which takes into account the relative 
contribution of each source of non-true $b$-jets is derived. The average mistag rate $\epsilon_f$ is
defined in terms of the various jet-flavour efficiencies as 
$\epsilon_f = f_\tau\epsilon_\tau+f_c\epsilon_c+f_{\mathrm{l}}\epsilon_{\mathrm{l}}$, where 
$f_\tau,f_c$ and $f_{\mathrm{l}}$ are the relative fractions of each jet-flavour prior to any $b$-tagging requirement.
Since the reducible background is dominated by \ttbar\ events, the relative jet-flavour fractions are extracted from the  
\ttbar\ MC sample described in section~\ref{sec-samples},  separately for each lepton multiplicity. In 
events containing zero or one lepton, they are obtained as a function of the jet $\pt$ and $|\eta|$, 
and as a function of the jet multiplicity to take into account the dependence with the number of additional partons produced in the hard-scattering or in the radiations.    
In events with exactly one lepton, the contribution from hadronic $\tau$-jets arising from the second $W$ boson decay increases in events with $\mt > m(W)$ and therefore the relative fractions of each jet-flavour are additionally binned as a function of the transverse mass. Events with two leptons are present in the inclusive 1-lepton SRs and in a 2-lepton control region (CR) used for the determination of the dominant irreducible background contribution. This CR is obtained by requiring $\mt < 140$~\gev\ to prevent overlap with the 1-lepton SRs as detailed in section~\ref{sec-fit}. In dileptonic \ttbar\ events, both $W$ bosons decay into an electron, a muon or a leptonically decaying $\tau$ and mistagged $b$-jets can only come from additional  $c$- or light-parton-jets. Consequently,  the jet-flavour fractions are only parameterised as a function of the jet \pt\ and $\eta$ in events with two leptons. 

Alternatively, the average mistag rates are determined in data 
using 0, 1 and 2-lepton regions enriched in $\ttbar$ events. These regions are defined 
following the same requirements as for the baseline event selection for all channels, except that events are required 
to have at least two $b$-jets and \met\ between 100~\gev\ and 200~\gev\ in order to minimise any possible 
contribution from signal events in the data. To estimate the mistag rate, the contribution from events with at least three true $b$-jets 
is subtracted using MC simulations. The mistag rate is measured 
as the probability to have a third $b$-jet in bins of $\pt$ and $|\eta|$, and an additional parameterisation as a function of $\mt$ is used in the 1-lepton channel. 
The results obtained with this method are consistent with the ones based on MC simulations. 
Because of the low number of events in data, the mistag rate estimated from MC simulations using the jet-flavour fractions 
method is taken as baseline, and the difference with the measurement in data 
is treated as a systematic uncertainty. 

This procedure was validated using the inclusive sample of simulated \ttbar\ events 
described in section~\ref{sec-samples} as follows. The MM is applied to the entire MC sample to predict the 
number of events with at least one mistagged $b$-jet. The contribution from the irreducible \ttbb\ background  
is extracted from the same sample as detailed in section~\ref{sec-samples}, and the sum of the two components is compared to the 
 inclusive event yield of the MC sample. Good agreement is found, at preselection level and also at various steps in the event selection chain.

\subsection{Irreducible background} \label{sec-fit}

The estimate of the minor contribution from $\ttbar+Z/h$ production followed by $Z/h\rightarrow b\bar{b}$ 
relies on MC predictions normalised to their theoretical cross-sections, while the dominant 
irreducible background from \ttbb\ events is estimated by normalising the MC predictions 
to the observed data in a CR. 
The CR, common to all SRs in both the 0- and 1-lepton channels, is defined using events 
with exactly two signal leptons and at least four jets with $\pt>30$~\GeV, at least one of them being required to have $\pt>90$~\GeV\ and  three of them to be  
$b$-tagged. The \met\ threshold is relaxed to 100~\GeV\ to increase the sample size, and  the transverse mass is required to be less than 140~\GeV\ to remove the overlap  with the 1-lepton SRs and to reduce the potential contamination from signal events to below a few per cent. The trigger efficiency is above 90\% in the CR and a systematic uncertainty of 2\% is added to
account for a small difference between the trigger turn-on curves in data and MC simulations in the 100--150~\GeV\ \met\ range. 
 Figure~\ref{f-CR}
shows the \mt distribution in the CR, before the requirement of $\mt < 140$~\gev;   the jet multiplicity, the \met, and the \meffi distributions with $\mt < 140$~\GeV\ are also shown. 

\begin{figure*}[htp]
\centering
\subfigure[]{\includegraphics[width=0.49\columnwidth]{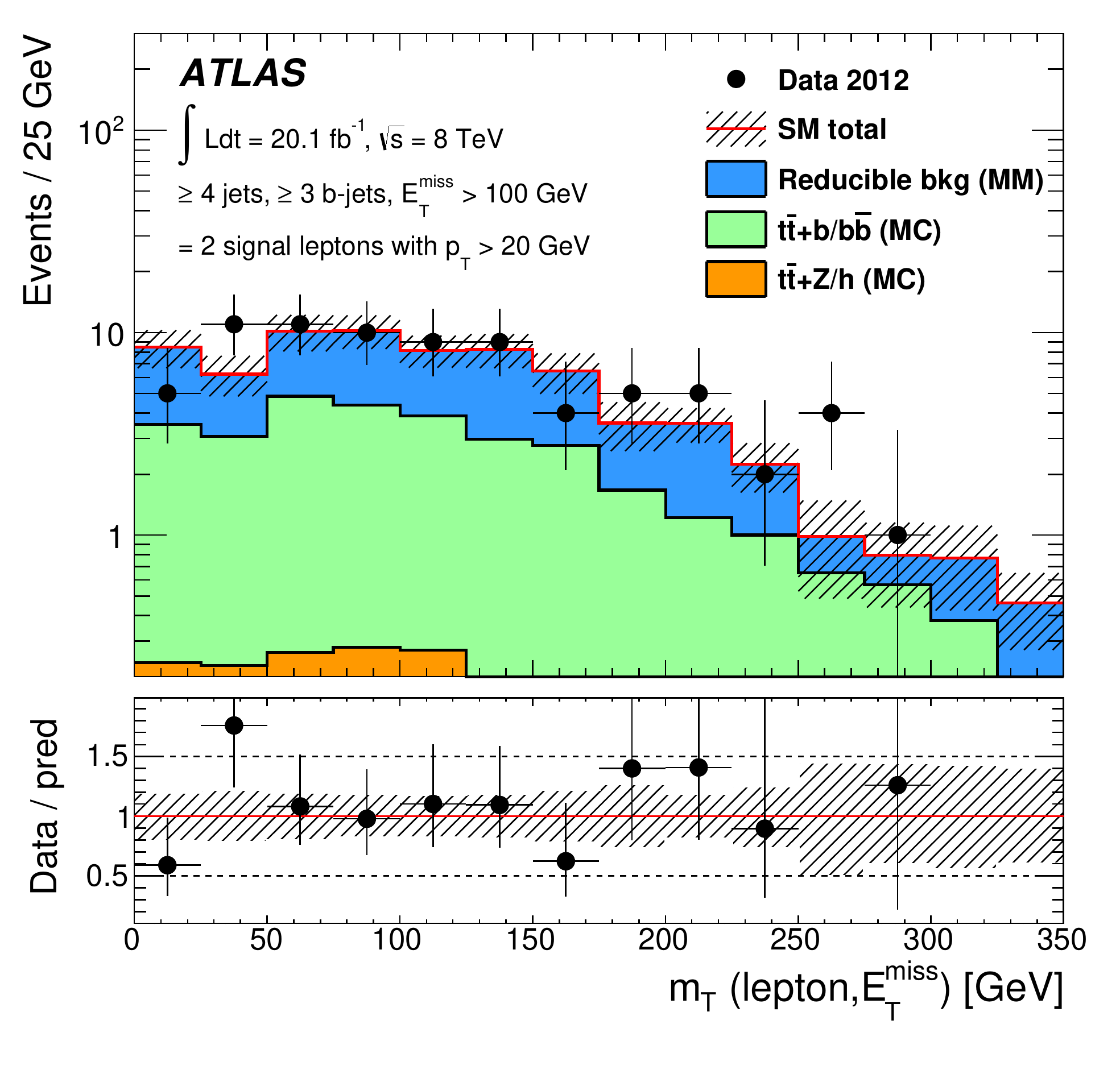}}
\subfigure[]{\includegraphics[width=0.49\columnwidth]{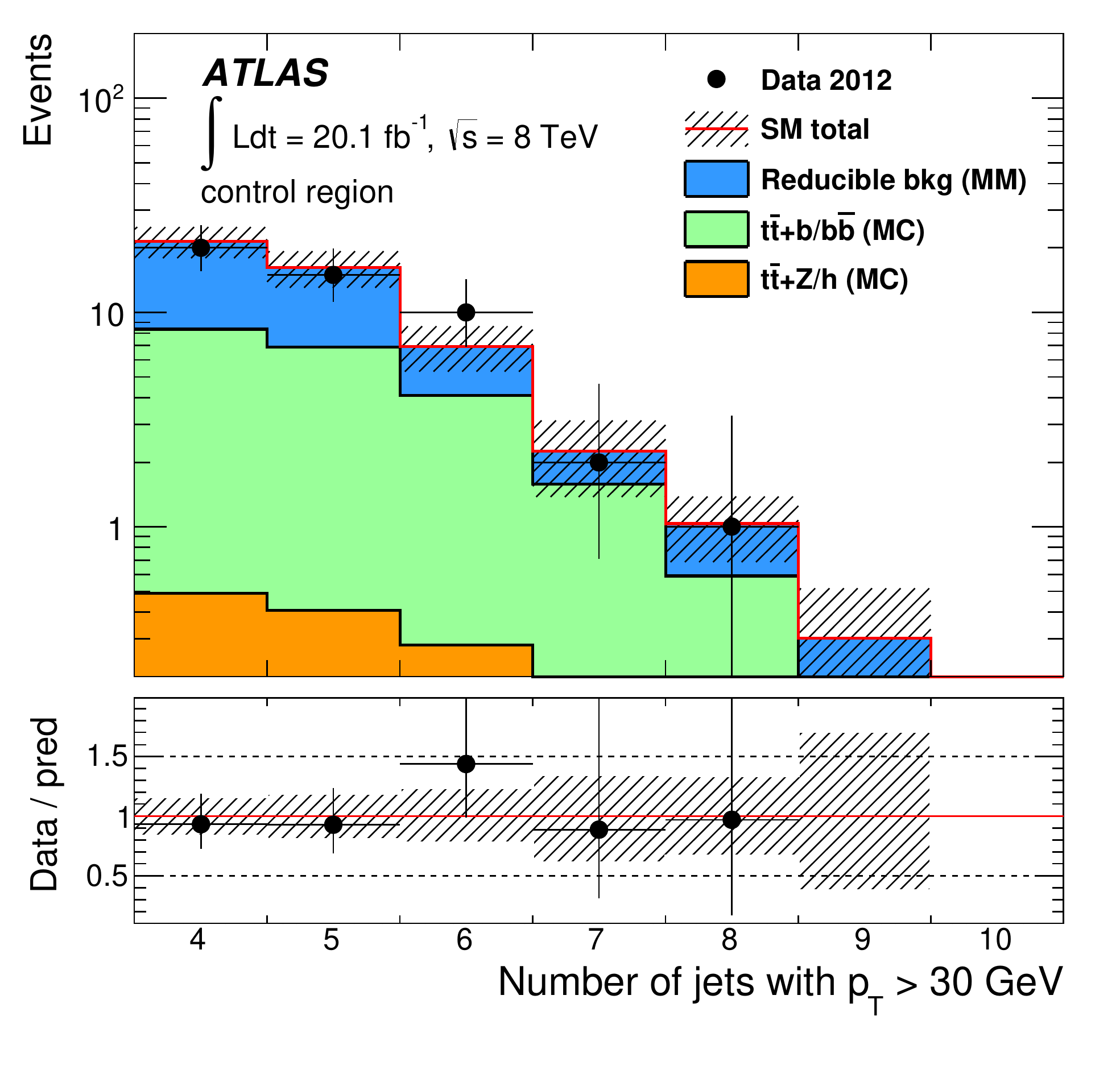}}\\
\subfigure[]{\includegraphics[width=0.49\columnwidth]{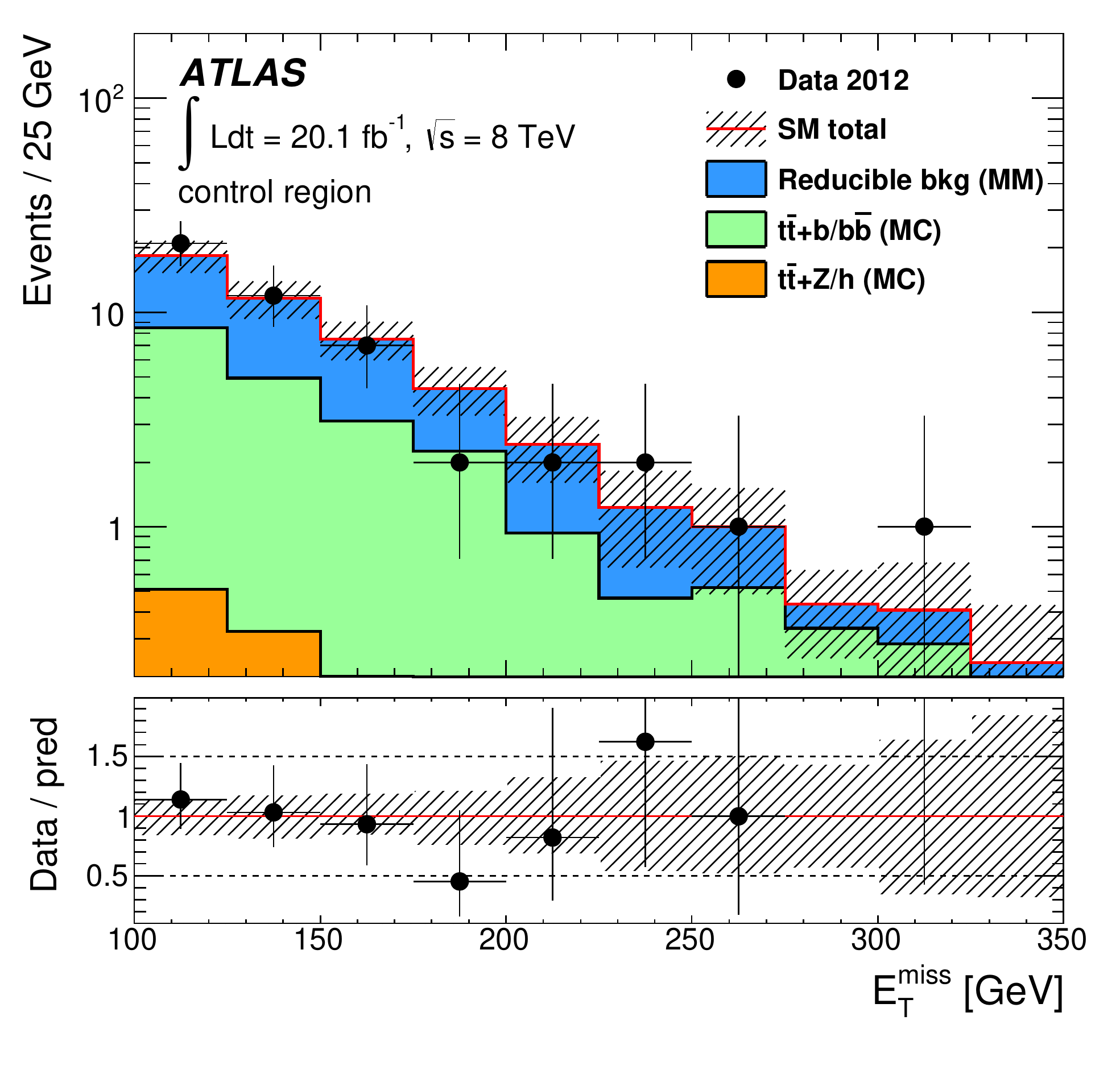}} 
\subfigure[]{\includegraphics[width=0.49\columnwidth]{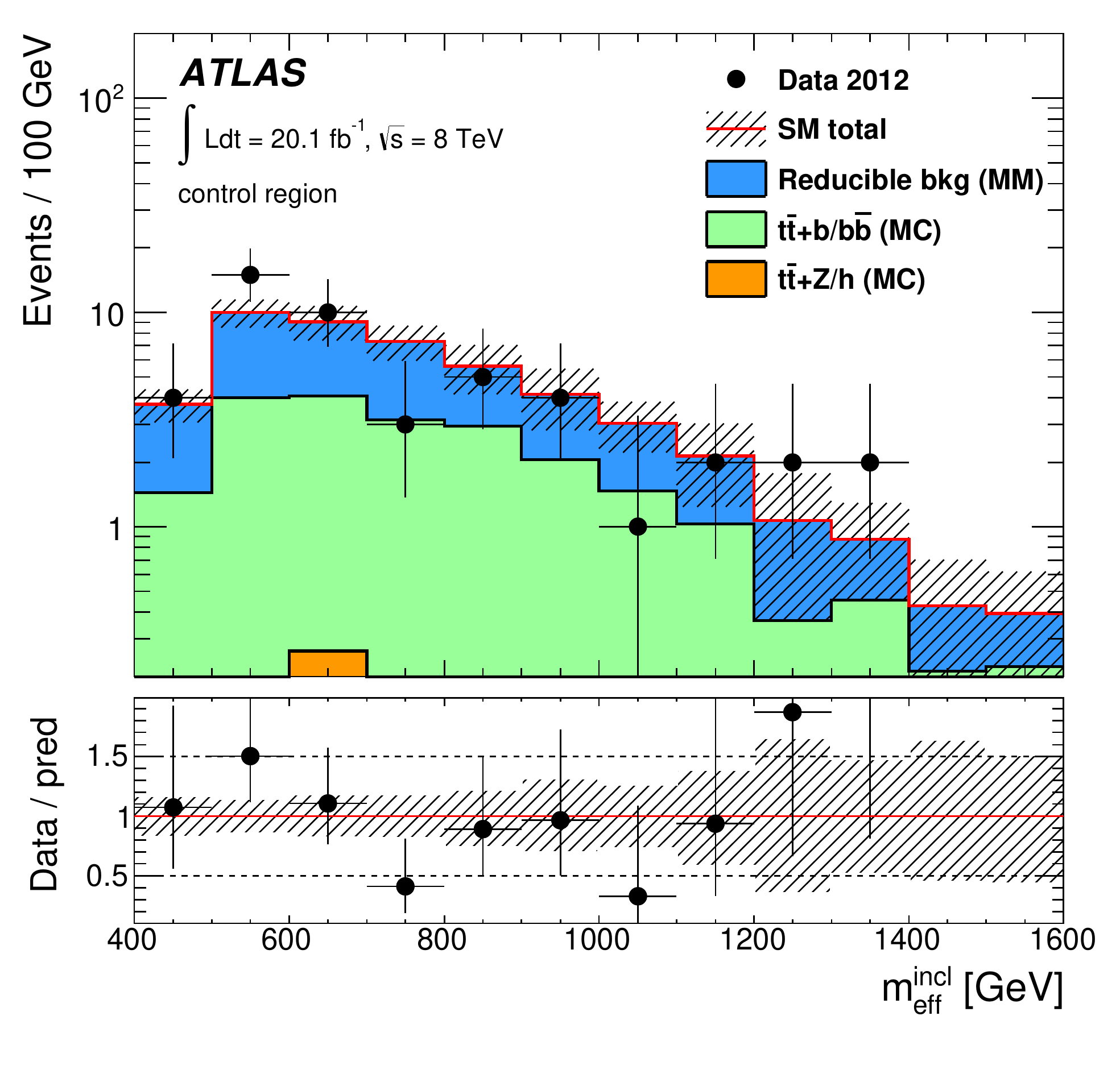}}
\caption{Expected distributions of SM background events and observed data distributions in the 2-lepton control 
region. The distributions of (a) \mt\  prior to the requirement on this variable, and (b) the number of 
jets with \pt\ $>$ 30 \gev, (c) \met\ and (d) \meffi\ are shown.   
Also displayed are the respective contributions of the backgrounds described in the legend   
 and the ratio between the expected and observed event yields.  
 The shaded bands include all experimental systematic uncertainties on the MC and  MM predictions. 
 The normalisation of the irreducible background \ttbb\ is as predicted by its theoretical
  cross-section scaled to the same luminosity as the data, prior to the fit in the control region. 
}\label{f-CR}
\end{figure*}

The expected number of \ttbb\ events in the various SRs is estimated
via a profile likelihood fit~\cite{profilelike} to the events in the 2-lepton CR. 
The expected and observed numbers of events in the CR are described  by Poisson probability functions. 
The statistical and systematic uncertainties  on the expected values described in section~\ref{sec-sys} are treated as nuisance parameters and are constrained by a Gaussian function 
with a width corresponding to the size of the uncertainty considered, taking into account the correlations between these parameters.  
The likelihood function is built as the product of the Poisson probability functions and the constraints on the nuisance parameters. 
The free parameter is the  overall normalisation of the \ttbb\ background, 
while the normalisations of the remaining irreducible and reducible backgrounds are initially set to their predictions and 
allowed to vary within their systematic uncertainties.  
The result of the fit in the CR is summarised in table~\ref{t-bkgfit}. 
Given the good agreement between the expected and observed yields,  
the fit gives a negligible correction to the normalisation of the \ttbb\ background. The uncertainty on the total background estimate is smaller 
than the largest individual uncertainty due to anticorrelations between the uncertainties on 
the reducible and irreducible backgrounds.

\begin{table*}[htp]
\centering
\begin{tabular}{lcc}
\toprule
 
 & Before the fit & After the fit \\
\midrule

Observed events   &  48   & 48 \\
\midrule
Total background events &  $48$ &   $48 \pm 7$ \\
\midrule
Reducible background events & $27$ &   $27 \pm 7$ \\
\ttbb\ events & $20$ &    $20 \pm 10$ \\
\ttbar+($Z\rightarrow b\bar{b}$) events & $0.5$ & $0.5 \pm 0.2$  \\
\ttbar+($h \rightarrow b\bar{b}$) events & $0.9$ & $0.9 \pm 0.9$  \\
\midrule
MC-only prediction & 50 & - \\
\bottomrule
    \end{tabular}
\caption{Background fit result in the  \ttbb\ CR. 
Uncertainties quoted include statistical and detector-related systematic effects. 
The MC-only prediction is given for comparison. 
}
\label{t-bkgfit}
\end{table*}

\section{Systematic uncertainties}\label{sec-sys}

The dominant detector-related systematic uncertainties on the amount of irreducible background are due to the jet energy scale
(JES) and resolution (JER) uncertainties; they range respectively from 16\% to 37\% and from 1\% to 32\% in the various SRs 
before the fit.
The JES uncertainty is derived from a combination of simulations, test beam data
and {\it in situ} measurements~\cite{JES,JES2}. Additional contributions accounting for the
jet-flavour composition, the calorimeter response to
different jet-flavours, pile-up and $b$-jet calibration uncertainties are also taken into account.
 Uncertainties on the JER are obtained with an {\it in situ} measurement of the
jet response asymmetry in dijet events. Uncertainties in jet measurements are
propagated to the \met, and additional subdominant uncertainties on \met\  arising from
energy deposits not associated with any reconstructed objects are also included.
The uncertainty associated with $b$-jets is evaluated by varying the $\pt$- and flavour-dependent
correction factors applied to each jet in the simulation within a range that reflects the systematic
uncertainty on the measured tagging efficiencies and mistag rates. 
It varies between 10\% and 16\% in the different SRs for the irreducible background, 
but it largely cancels for the \ttbb\ background because of the normalisation in the CR. 
Uncertainties in lepton reconstruction and momentum scales are negligible. 
All these experimental systematic uncertainties are treated as fully correlated between the signal
and the irreducible backgrounds extracted from MC simulations.

Additional theoretical systematic uncertainties are considered for the irreducible backgrounds. 
The uncertainty on the \ttbb\ cross-section cancels in the normalisation of the MC simulation 
in the CR, and only the theoretical uncertainties on the MC prediction used to extrapolate from the CR  to the 
SRs are considered.  The uncertainty due to the choice of the 
factorisation ($\mu_{\mathrm{F}}$) and renormalisation ($\mu_{\mathrm{R}}$) scales in {\textsc POWHEG} are estimated 
by comparing the baseline sample to  {\textsc POWHEG}+{\textsc PYTHIA-6} samples generated with 
$\mu_{\mathrm{F}}$ and $\mu_{\mathrm{R}}$ varied separately up and down by a factor of two, giving an
uncertainty of up to 13\%. 
The uncertainty due to the choice of  MC generator is estimated by comparing the estimate from the 
nominal \mbox{{\textsc POWHEG}+{\textsc PYTHIA-6}}  sample 
to the  {\textsc MadGraph}+{\textsc PYTHIA-6} sample generated  with up to three additional partons at the matrix-element  
level, yielding an uncertainty of up to 30\%. The parton shower uncertainty is assessed by comparing
{\textsc POWHEG} interfaced to {\textsc PYTHIA-6} with  {\textsc POWHEG} interfaced
to {\textsc HERWIG} and {\textsc JIMMY}, and amounts up to 65\% in the SRs where at least seven jets are required. 
The PDF uncertainties are derived following the Hessian method~\cite{Pumplin:2001ct}, resulting in an uncertainty of less than 2\% for all SRs.  

The theoretical uncertainty on the $\ttbar+Z$ cross-section is 22\%~\cite{Garzelli:2012bn}. 
 The systematic uncertainties associated with the modelling of $\ttbar+Z$ events are estimated by using different 
{\textsc MadGraph}+{\textsc PYTHIA-6}  samples: 
variations up and down of the $\mu_{\mathrm{R}}$ and $\mu_{\mathrm{F}}$ by a factor of two result in an uncertainty
of up to 50\%; variations of the ISR/FSR parameters within ranges 
validated by measurements in data yield an uncertainty of up to 50\% for each variation;    
variations up and down of the matrix-element to parton-shower matching parameter \textit{xqcut} by 5~\GeV\ around the central value of 20~\gev\ result in an uncertainty of up to 30\%. For the small contribution from the 
$\ttbar+h(\rightarrow b\bar{b})$ background, a total uncertainty of 100\% is assumed to account for large uncertainties on the acceptance, 
while the inclusive cross-section is known to better precision. 

Systematic uncertainties on the MM prediction of the reducible backgrounds include 
the uncertainties on the measurement of the $b$-tagging efficiency for the different jet-flavours. They vary in the range between 4\% and 14\% and are treated as fully 
correlated with the irreducible background and the signal. The statistical uncertainty of  the \ttbar\ MC sample  
used to extract the jet-flavour fractions is also taken into account and is of the order of 1\%.
 The difference between the baseline prediction obtained with the mistag rate from simulated \ttbar\ events  
and the prediction obtained using the mistag rate measured in data is assigned as a systematic
uncertainty. This uncertainty ranges from 9\% to 45\% in the different SRs. 
Finally, the statistical uncertainty  on the number of observed events for each $b$-jet multiplicity is
propagated to the MM prediction. This latter uncertainty is the dominant source of uncertainty on the background estimation in most SRs.

\section{Results} \label{sec-res}

The data are compared to the background predictions in figures~\ref{fig:presel1_0l}--\ref{fig:SR_1l}.
Figure~\ref{fig:presel1_0l} shows the observed distributions of the  number of jets with $\pt >$ 30 \gev, \meffe\ and \met\  
 after the 0-lepton baseline selection detailed in table~\ref{t-SR}, together with 
the background prediction from the MM for the reducible background and from MC simulations for the irreducible 
background. 
Figure~\ref{fig:presel2_0l} shows the same distributions for events with at least four jets with $\pt >$ 50~\gev\ and three $b$-jets with $\pt >$ 50~\gev\ after 
the 0-lepton baseline selection. 
The \meffi\ and \met\ distributions with a requirement of at least  
seven jets with $\pt >$ 30 \gev\ after the 0-lepton baseline selection are shown in figure~\ref{fig:presel3_0l}.    
Figure~\ref{fig:presel_1l} shows the number of jets with $\pt >$ 30 \gev\ after the 1-lepton baseline selection, as well as 
the distributions of \mt,  \met\ and \meffi\ after  requiring at least six jets with $\pt > 30$~\GeV\ in addition to the 1-lepton baseline selection.  
The \meffe, \meffi\  and \met\ distributions obtained before the final requirement on these quantities are shown in 
figures~\ref{fig:SR_0l_4j}--\ref{fig:SR_1l}, representing each SR.  Also shown in all figures are the predictions of two benchmark signal models. 

The background prediction in each SR is obtained by adding the Poisson probability function 
describing the expected number of events in the SR and the corresponding nuisance parameters in the likelihood fit.  
The results of the fits and the observed data in each SR are reported in tables~\ref{t-SR0L} and~\ref{t-SR1L} for the 0-lepton and 1-lepton channels, respectively.  
No significant deviation from the SM expectation is observed in any of the
0-lepton SRs. In the 1-lepton channel, a deficit in data is observed in all overlapping SRs. 
In addition to the event yields, the $CL_b$-values~\cite{cls}, which quantify the observed level of agreement with the expected yield, 
and the $p_0$-values, which represent the probability of the SM background alone to fluctuate to the observed number of events or higher, are also reported.
The $p_0$-values are truncated at 0.5 if the number of observed events is below the number of expected events.  
Upper limits at 95\%  confidence level (CL) on the number of beyond-the-SM (BSM) events  are derived in each SR using the $CL_s$ prescription~\cite{cls} 
and neglecting any possible signal contamination in the CR. 
These are obtained with a fit in each SR which proceeds in the same way as the fit used to predict the background, except that the number of events observed in the SR 
is added as an input to the fit, and an additional parameter for the non-SM signal strength, constrained to be non-negative, is fit. 
The upper limits are derived with pseudo-experiments, and the results obtained with an asymptotic approximation~\cite{profilelike} are given in parentheses for comparison. 
These limits, after being normalised by the integrated luminosity of the data sample, can be interpreted 
in terms of upper limits on the visible cross-section for hypothetical 
 BSM contributions,  defined in terms of the kinematic acceptance $A$ and the experimental efficiency $\epsilon$  
 as $\sigma_{\mathrm{vis}} = \sigma\times A \times\epsilon$. 

\begin{figure}[htp]
\centering
\subfigure[]{\includegraphics[width=0.49\columnwidth]{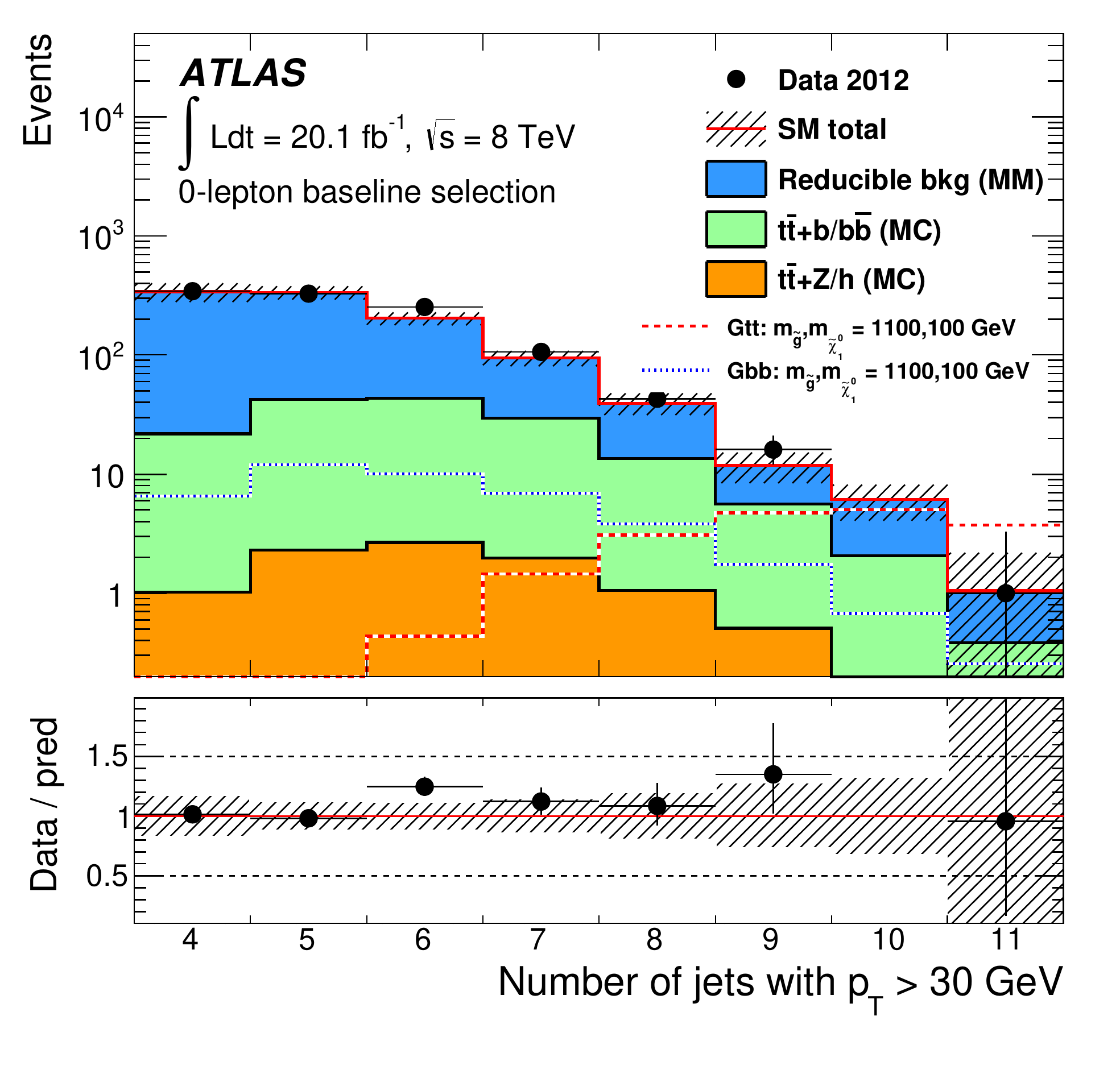}}  \\
\subfigure[]{\includegraphics[width=0.49\columnwidth]{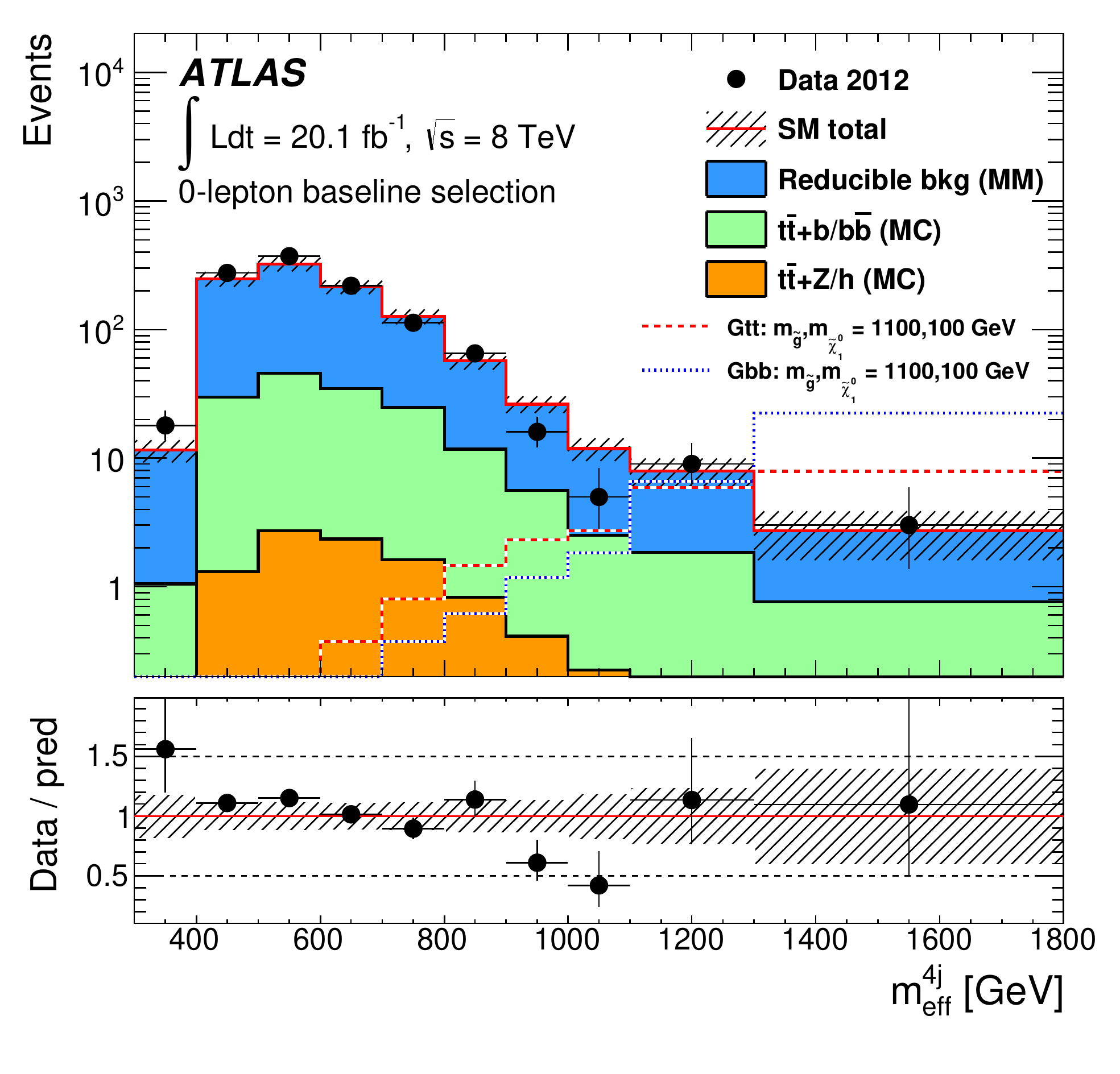}} 
\subfigure[]{\includegraphics[width=0.49\columnwidth]{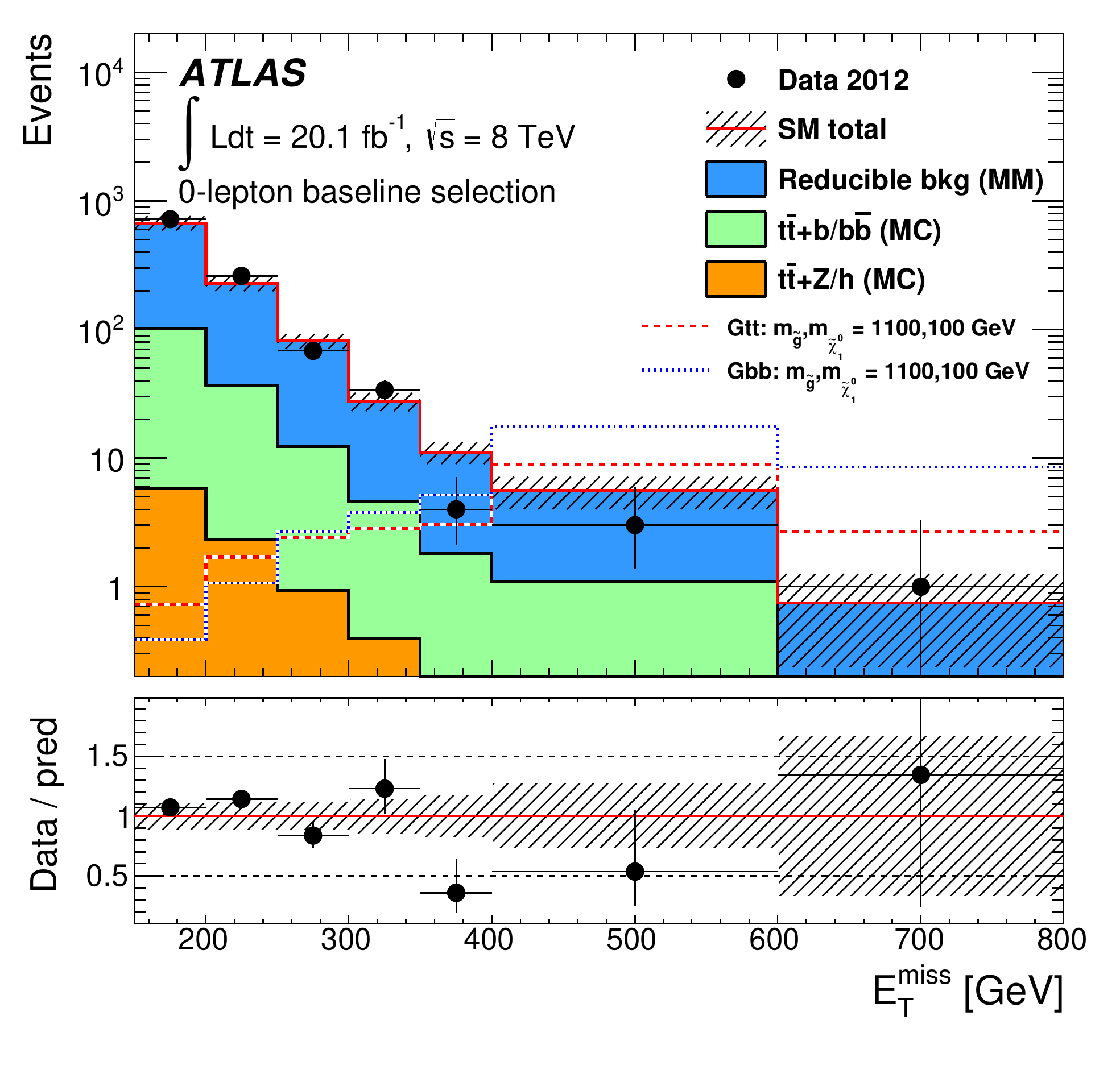}} \\ 
\caption{The observed distributions of (a) the number of jets with $\pt >$ 30 \gev, (b) \meffe\ and (c) \met\  
  after the 0-lepton baseline selection, together with the background prediction. 
Also displayed are the respective contributions of the backgrounds described in the legend   
 and the ratio between the expected and observed event yields.  
 The shaded bands include all  experimental systematic uncertainties on the background prediction.
The prediction for two signal points from the Gtt ($\gl \rightarrow t\bar{t} \tilde{\chi}_1^0$) and Gbb ($\gl \rightarrow b\bar{b} \tilde{\chi}_1^0$)   models are overlaid. 
 The normalisation of the irreducible background \ttbb\ is as predicted by its theoretical
  cross-section scaled to the same luminosity as the data, prior to the fit in the control region. } 
\label{fig:presel1_0l}
\end{figure}

\begin{figure}[htp]
\centering
\subfigure[]{\includegraphics[width=0.49\columnwidth]{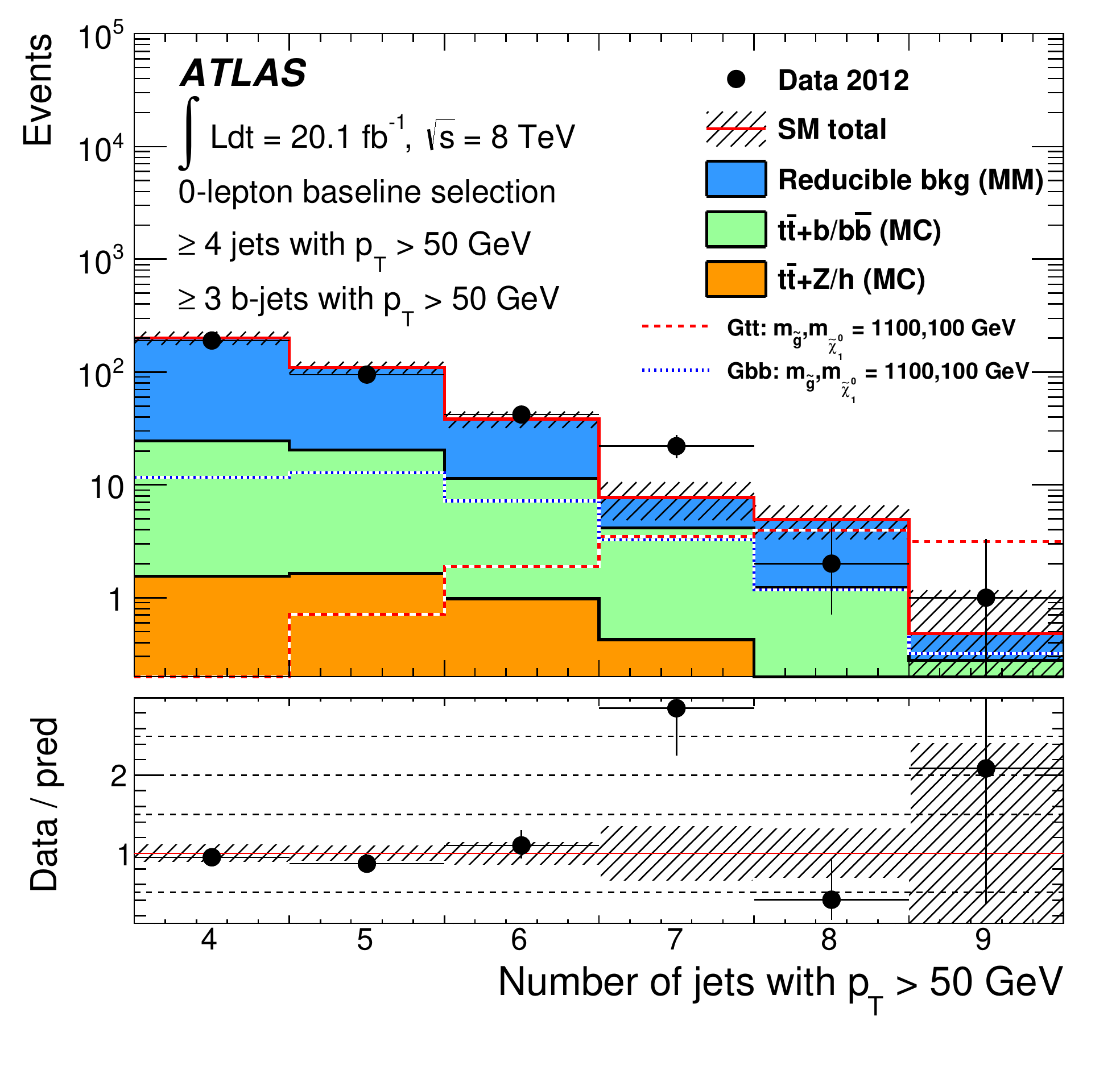}} \\ 
\subfigure[]{\includegraphics[width=0.49\columnwidth]{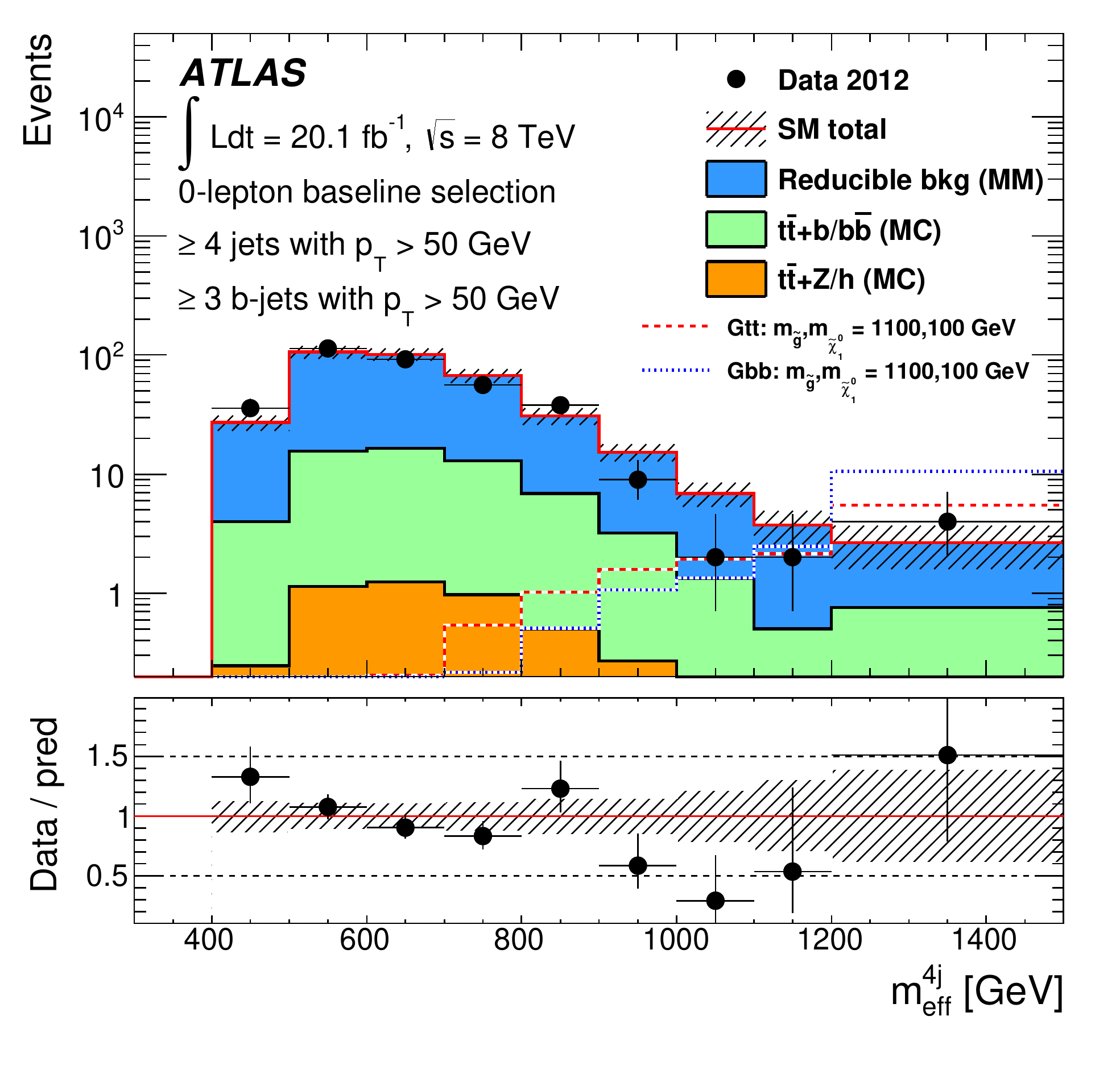}}
\subfigure[]{\includegraphics[width=0.49\columnwidth]{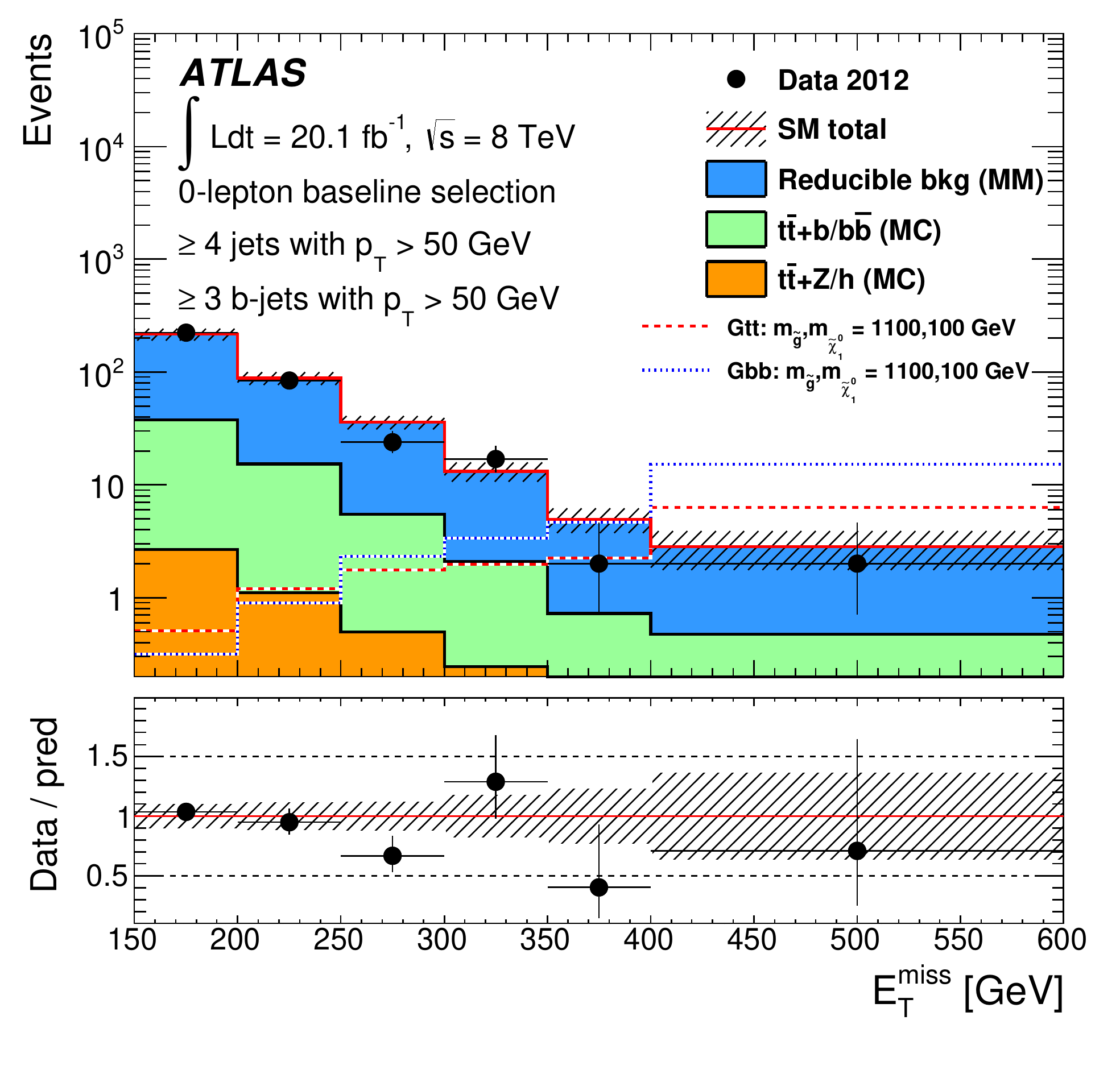}}\\ 
\caption{The observed distributions of (a) the number of jets with $\pt >$ 50 \gev, (b)  \meffe\   and (c) \met\     
  after requiring at least four jets with $\pt >$ 50 \gev\ and at least three $b$-jets with \mbox{$\pt >$ 50 \gev}  in addition to the 0-lepton baseline selection, 
 together with the background prediction.   
Also displayed are the respective contributions of the backgrounds described in the legend   
 and the ratio between the expected and observed event yields.  
 The shaded bands include all  experimental systematic uncertainties on the background prediction.
The prediction for two signal points from the Gtt \mbox{($\gl \rightarrow t\bar{t} \tilde{\chi}_1^0$)} and Gbb \mbox{($\gl \rightarrow b\bar{b} \tilde{\chi}_1^0$)}  models are overlaid.
 The normalisation of the irreducible background \ttbb\ is as predicted by its theoretical
  cross-section scaled to the same luminosity as the data, prior to the fit in the control region. } 
\label{fig:presel2_0l}
\end{figure}

\begin{figure}[htp]
\centering
\subfigure[]{\includegraphics[width=0.49\columnwidth]{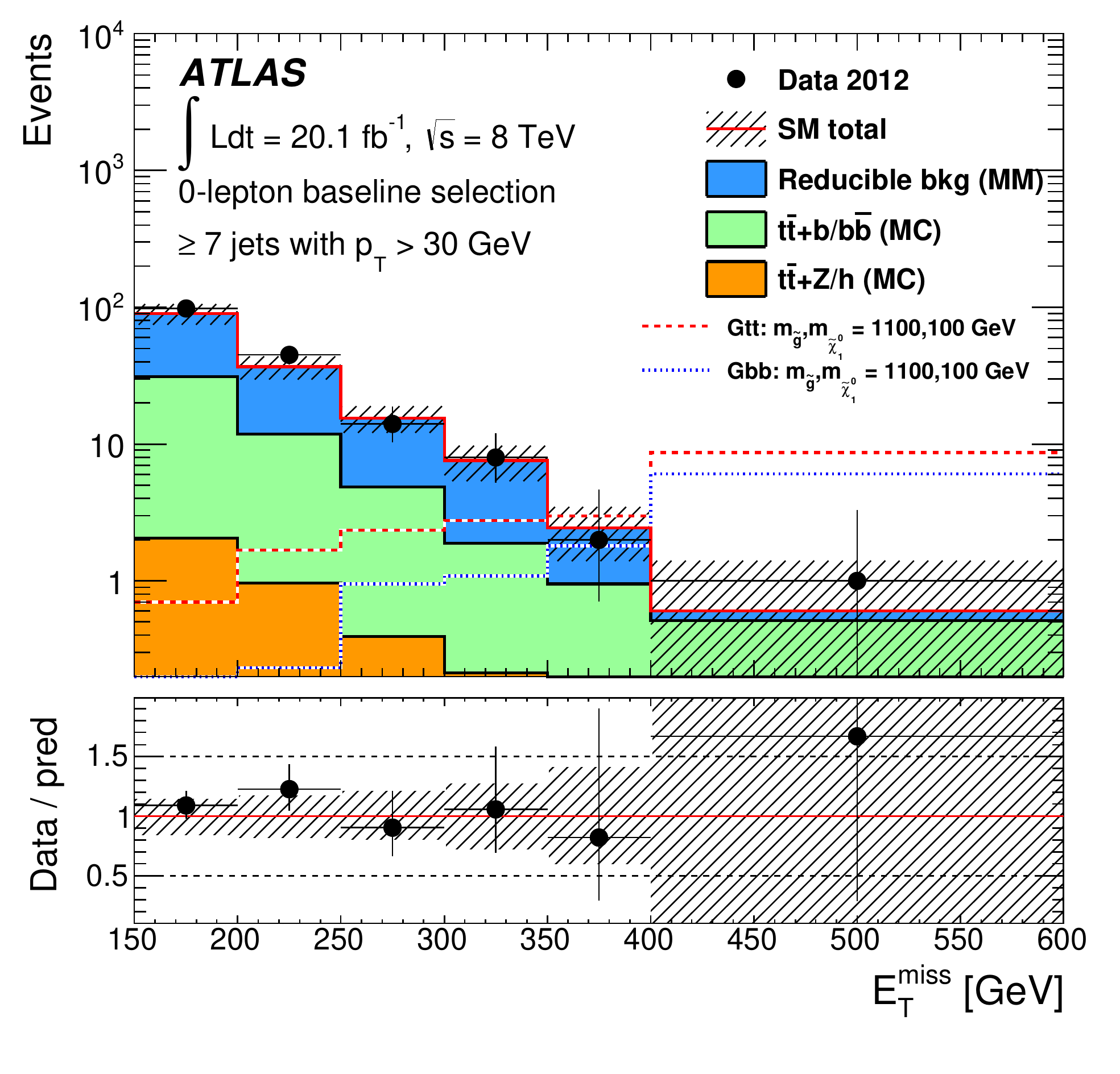}} 
\subfigure[]{\includegraphics[width=0.49\columnwidth]{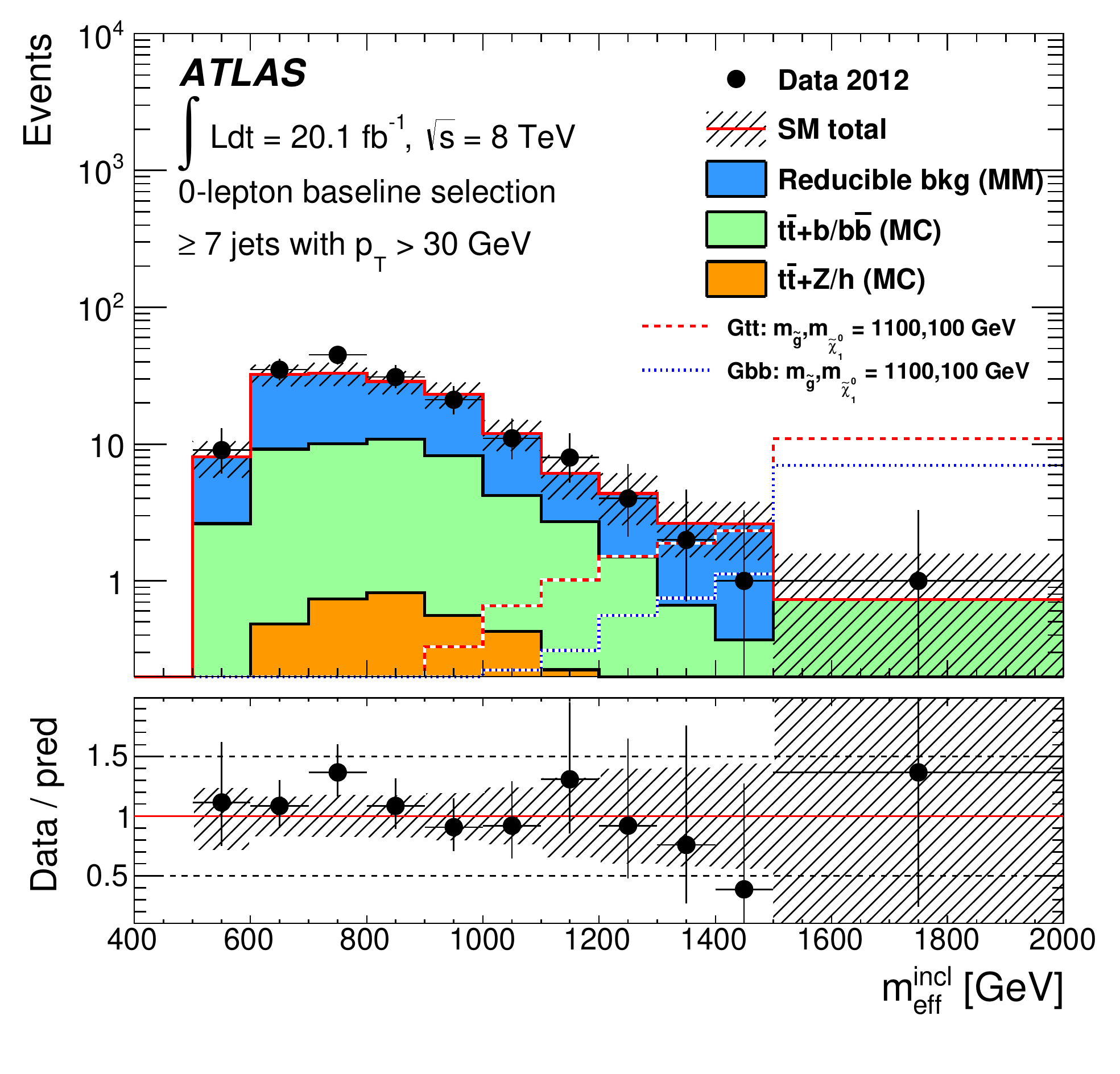}} \\ 
\caption{The (a) \met\  and (b) \meffi\  distributions  observed in data after requiring at least seven jets with $\pt >$ 30 \gev\ 
in addition to the 0-lepton baseline selection, together with the background prediction.  
Also displayed are the respective contributions of the backgrounds described in the legend   
 and the ratio between the expected and observed event yields. 
 The shaded bands include all  experimental systematic uncertainties on the background prediction.
The prediction for two signal points from the Gtt \mbox{($\gl \rightarrow t\bar{t} \tilde{\chi}_1^0$)} and Gbb \mbox{($\gl \rightarrow b\bar{b} \tilde{\chi}_1^0$)}   models are overlaid.  
The normalisation of the irreducible background \ttbb\ is as predicted by its theoretical
  cross-section scaled to the same luminosity as the data, prior to the fit in the control region. } 
\label{fig:presel3_0l}
\end{figure}

\clearpage 
\begin{figure}[htp]
\centering
\subfigure[]{\includegraphics[width=0.49\columnwidth]{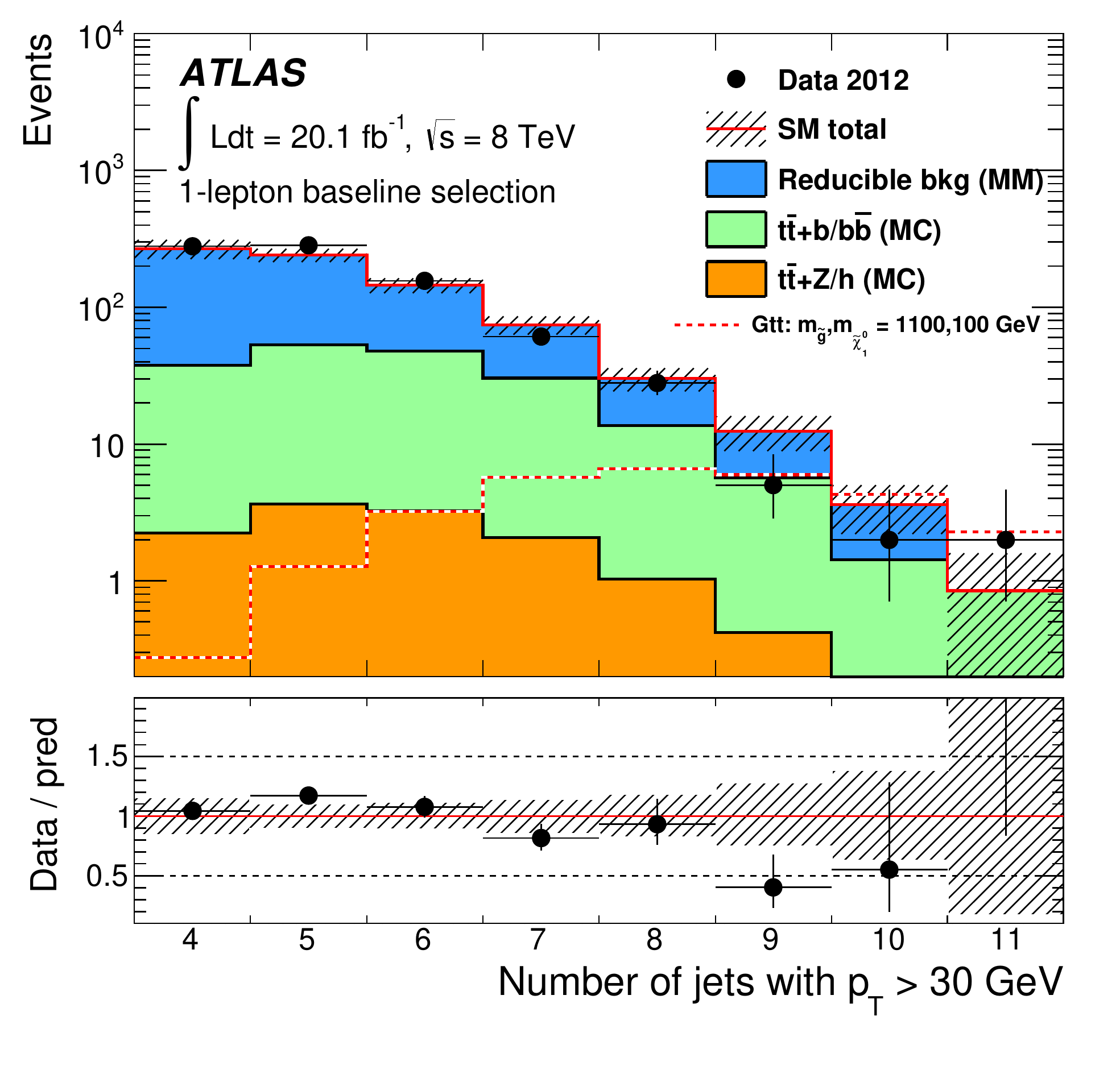}} 
\subfigure[]{\includegraphics[width=0.49\columnwidth]{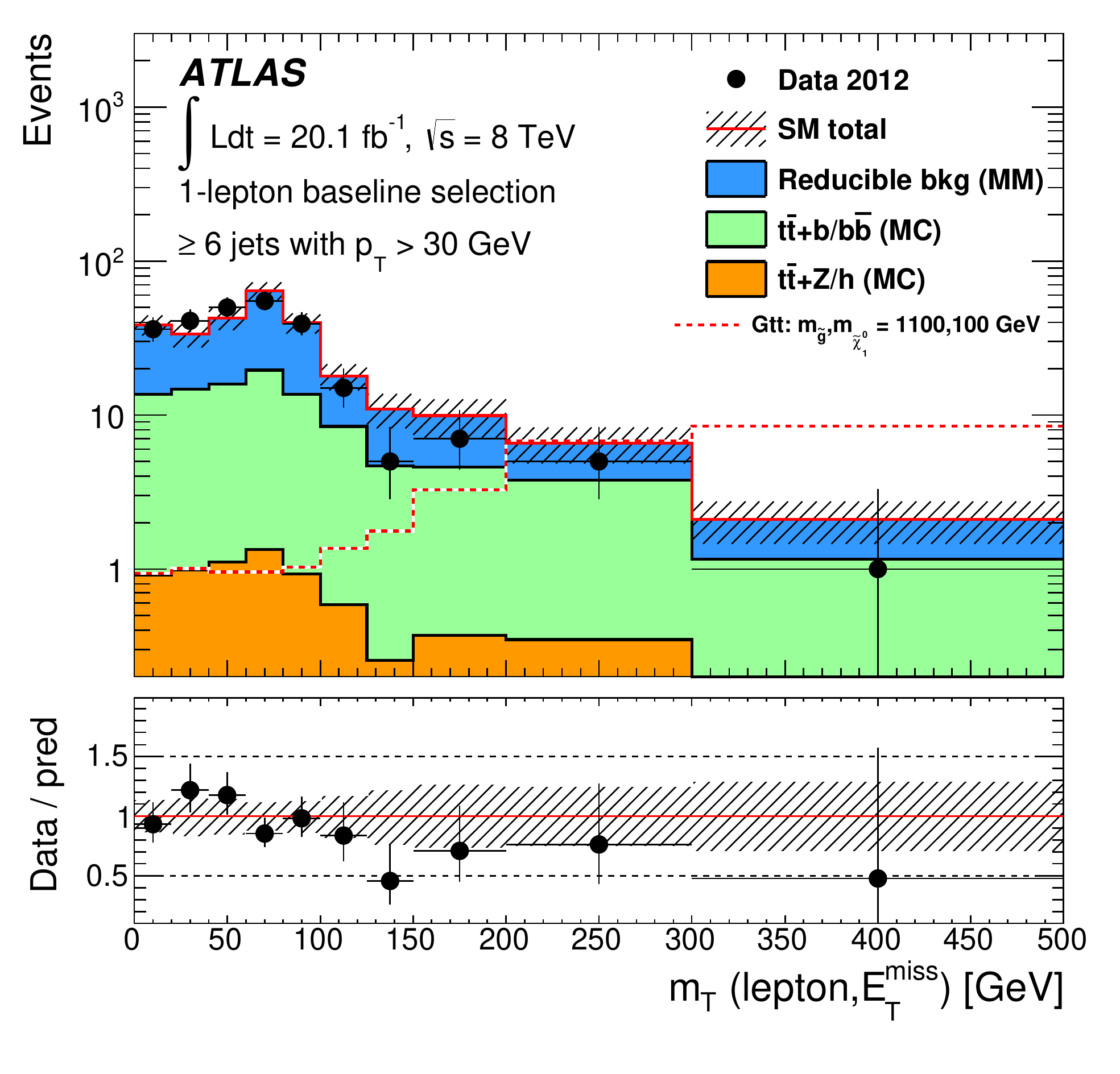}} \\
\subfigure[]{\includegraphics[width=0.49\columnwidth]{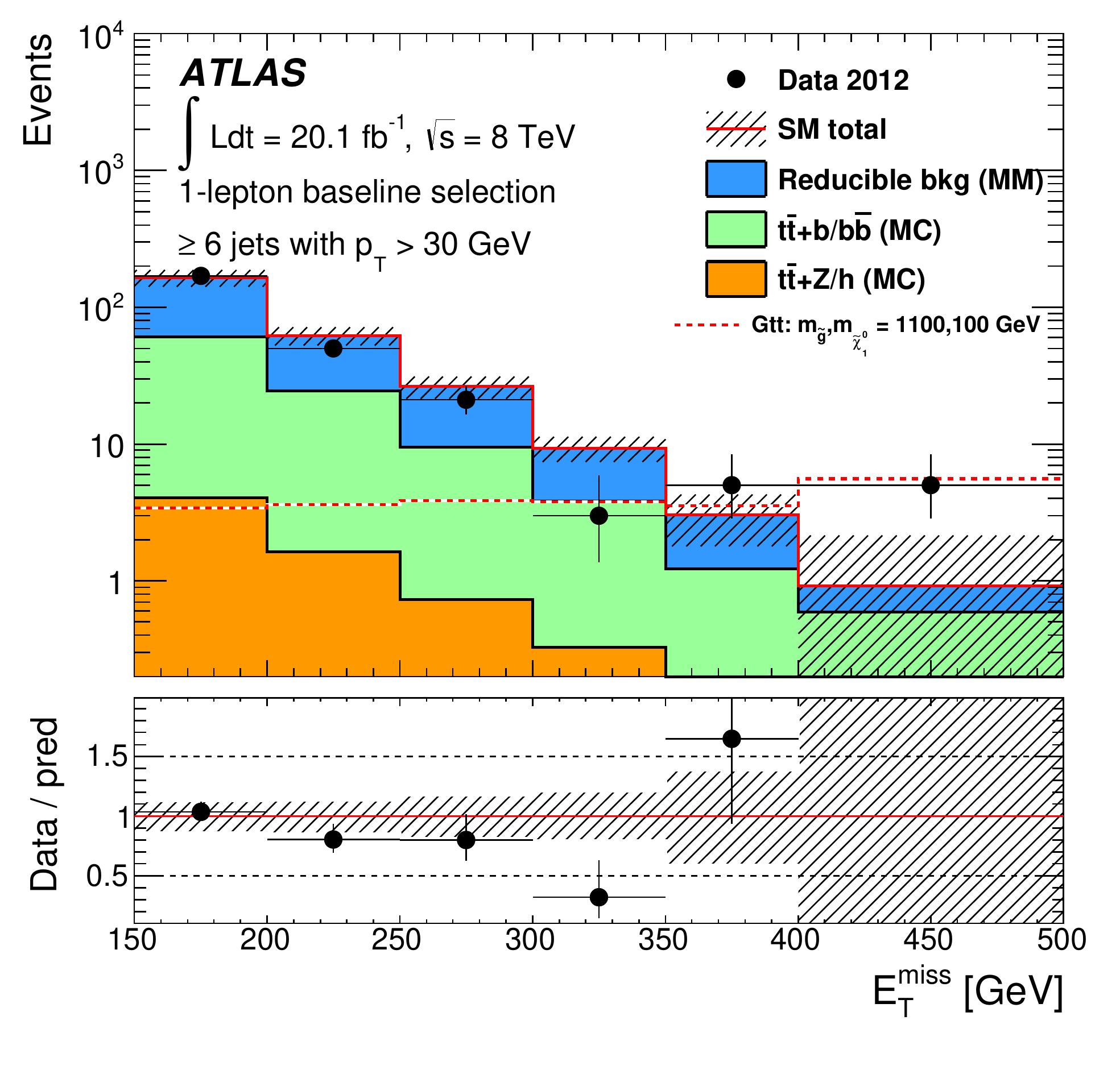}}
\subfigure[]{\includegraphics[width=0.49\columnwidth]{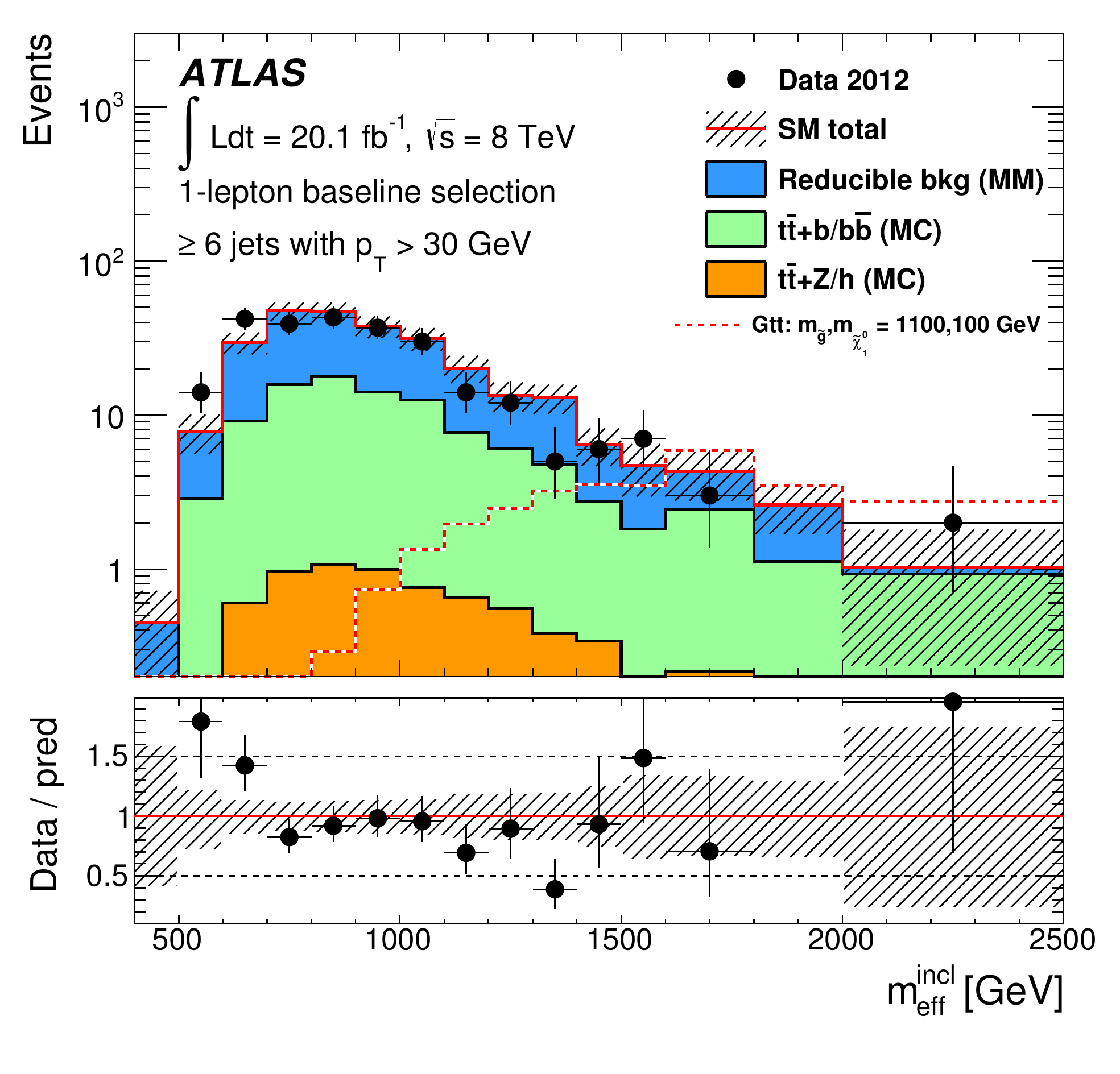}}  
\caption{The distribution of (a) the number of jets with $\pt >$ 30~\gev\ observed in data after the 
1-lepton baseline selection, together with the background prediction. The (b) \mt, (c) \met\  and (d) \meffi\   distributions after 
requiring at least six jets $\pt>30$~\GeV\ in addition to the 1-lepton baseline selection are also shown.
Also displayed are the respective contributions of the backgrounds described in the legend   
 and the ratio between the expected and observed event yields. 
 The shaded bands include all  experimental systematic uncertainties on the background prediction.
The prediction for one signal point from the Gtt ($\gl \rightarrow t\bar{t} \tilde{\chi}_1^0$)  model is overlaid. 
 The normalisation of the irreducible background \ttbb\ is as predicted by its theoretical
  cross-section scaled to the same luminosity as the data, prior to the fit in the control region. 
} 
\label{fig:presel_1l}
\end{figure}

\begin{figure}[htp]
\centering
\subfigure[]{\includegraphics[width=0.49\columnwidth]{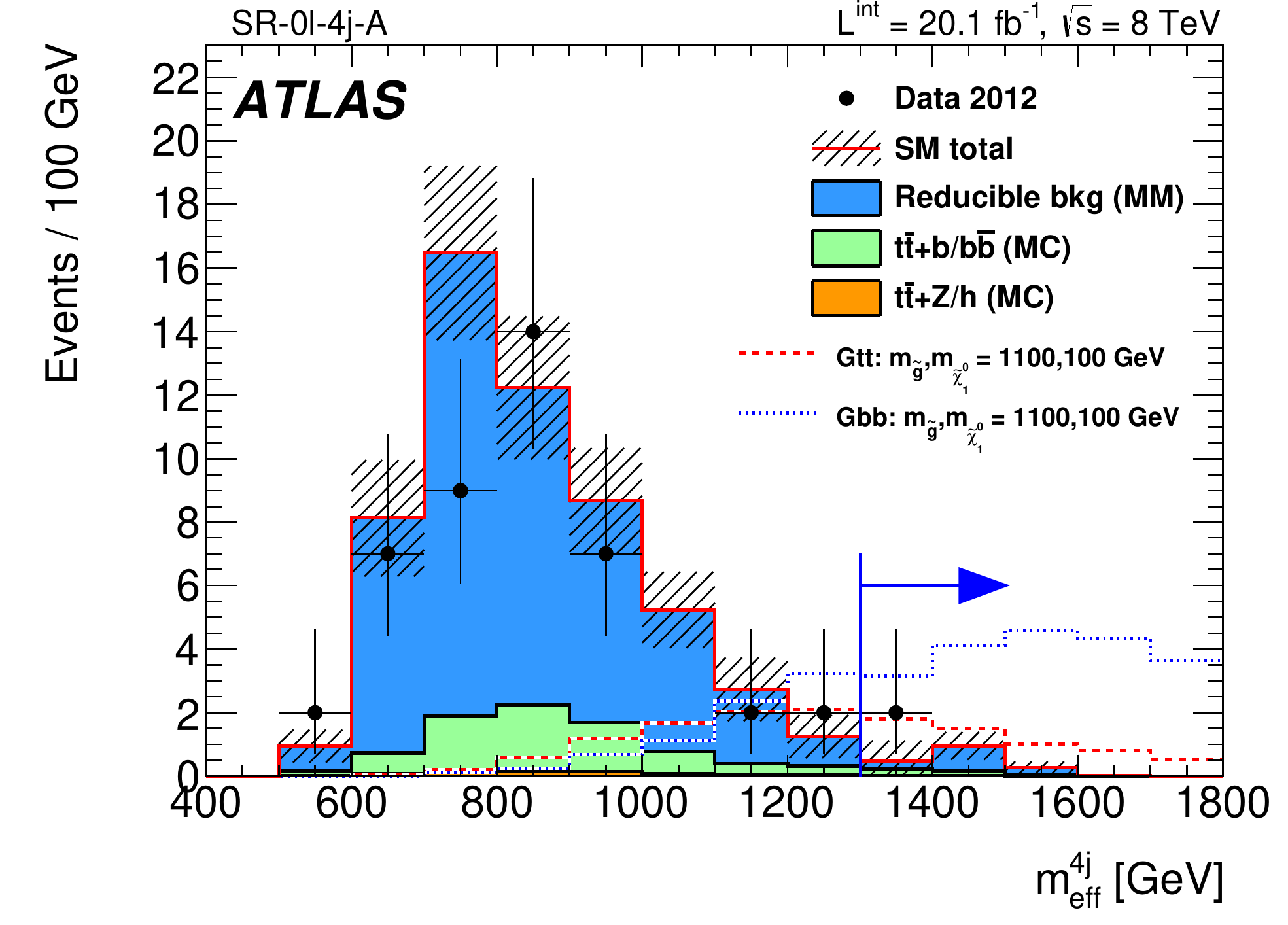}}
\subfigure[]{\includegraphics[width=0.49\columnwidth]{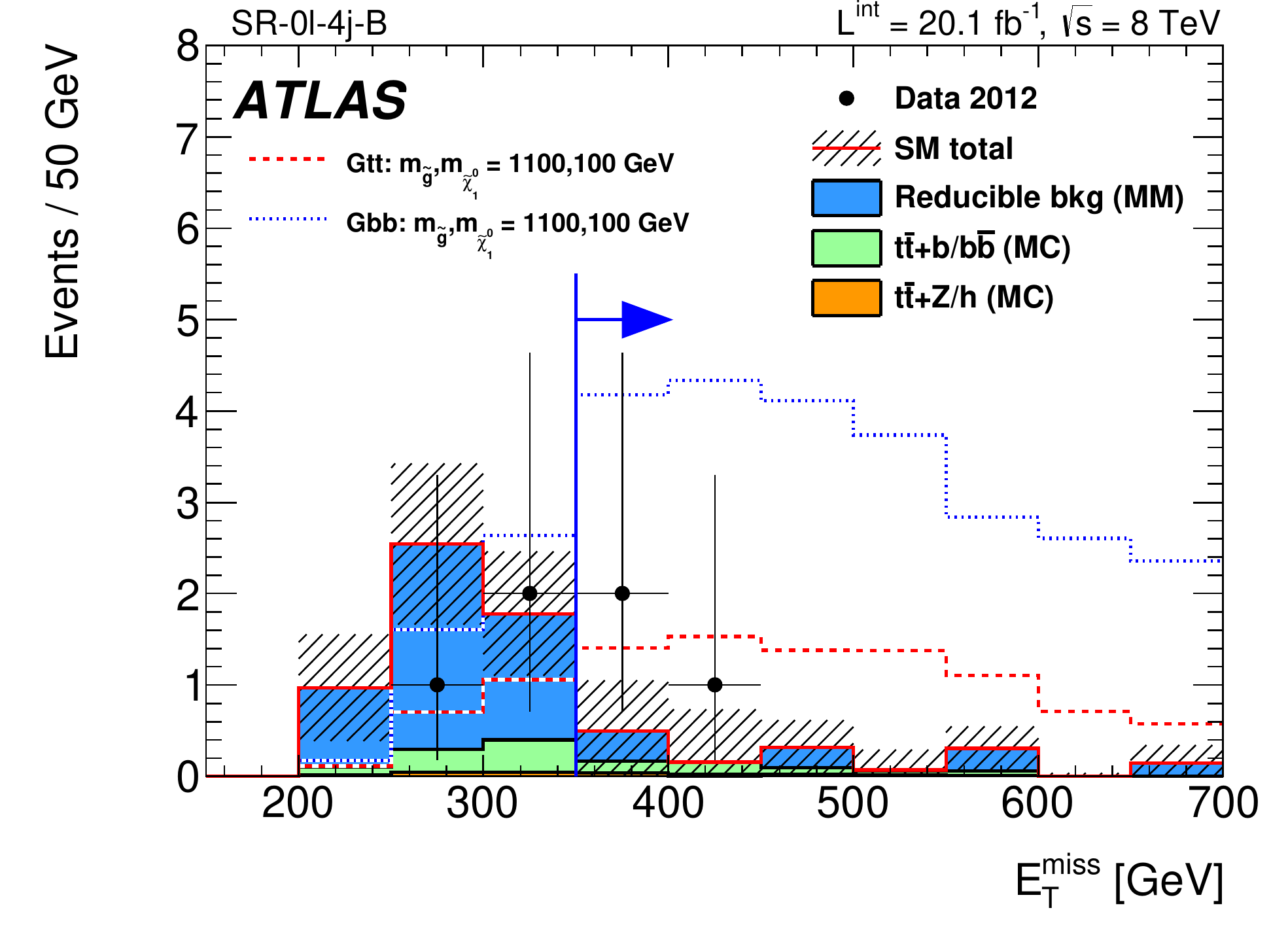}}\\ 
\subfigure[]{\includegraphics[width=0.49\columnwidth]{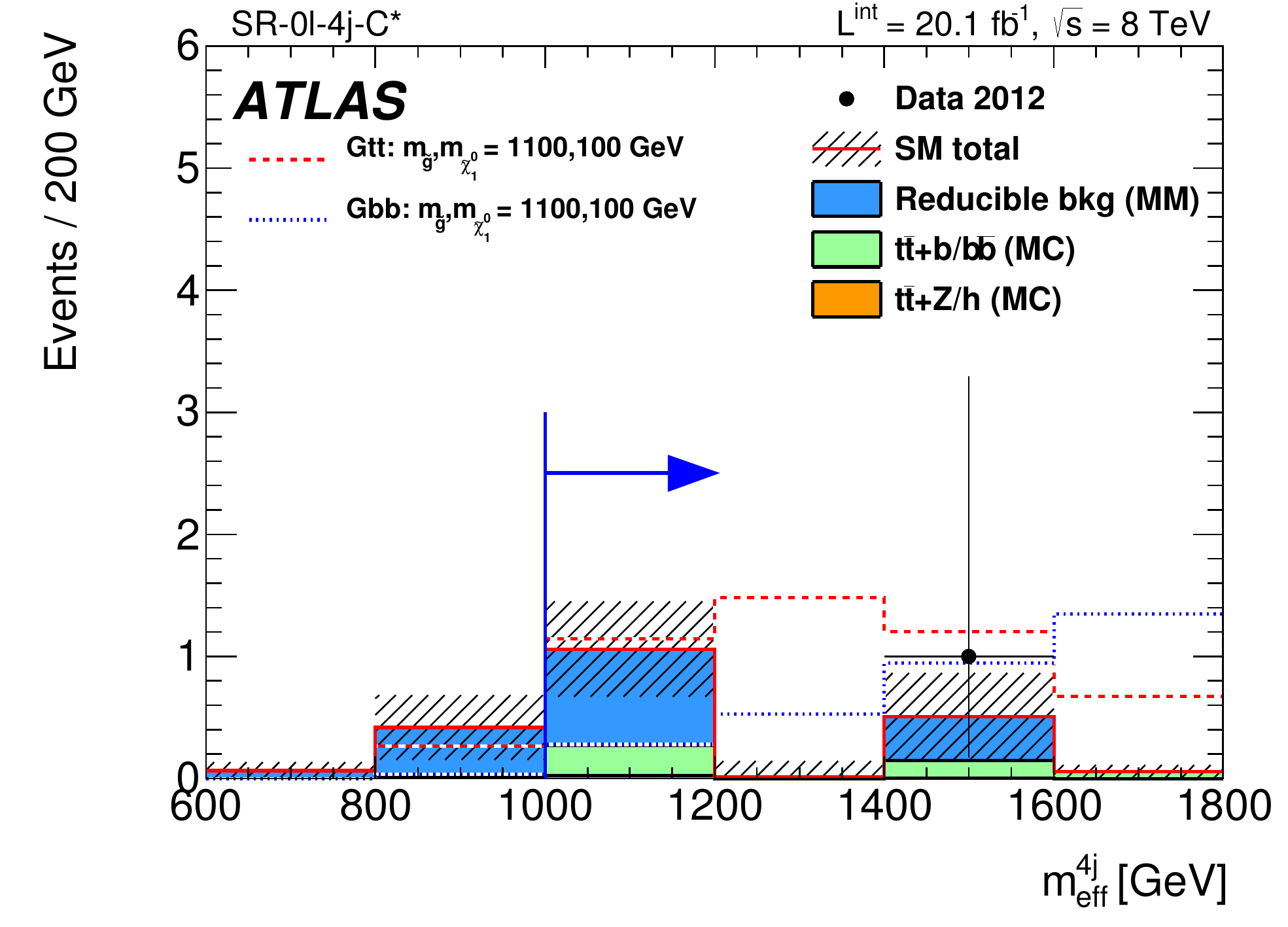}} \
\caption{The  (a)-(c) \meffe\  and (b) \met\  distributions observed in data for SR-0$\ell$-4j-A, SR-0$\ell$-4j-B and SR-0$\ell$-4j-C*, respectively,  
after all requirements applied but the one indicated by the arrow, together with the background prediction. 
 The shaded bands include all  experimental systematic uncertainties on the background prediction. 
The prediction for two signal points from the Gtt \mbox{($\gl \rightarrow t\bar{t} \tilde{\chi}_1^0$)} and Gbb \mbox{($\gl \rightarrow b\bar{b} \tilde{\chi}_1^0$)}   models are overlaid.  
 The normalisation of the irreducible background \ttbb\ is as predicted by its theoretical
  cross-section scaled to the same luminosity as the data, prior to the fit in the control region. 
} 
\label{fig:SR_0l_4j}
\end{figure}

\clearpage

\begin{figure}[htp]
\centering
\subfigure[]{\includegraphics[width=0.49\columnwidth]{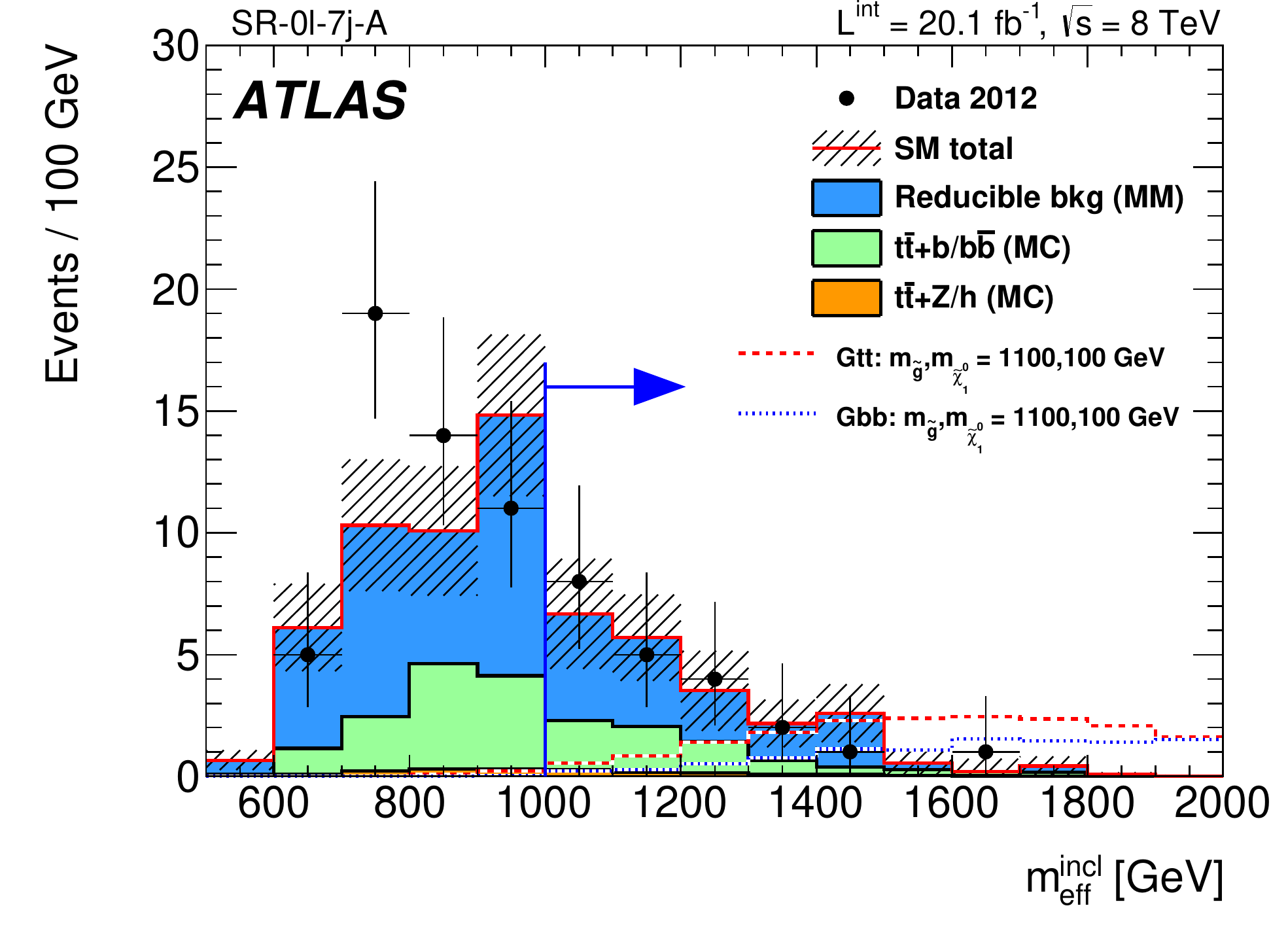}} 
\subfigure[]{\includegraphics[width=0.49\columnwidth]{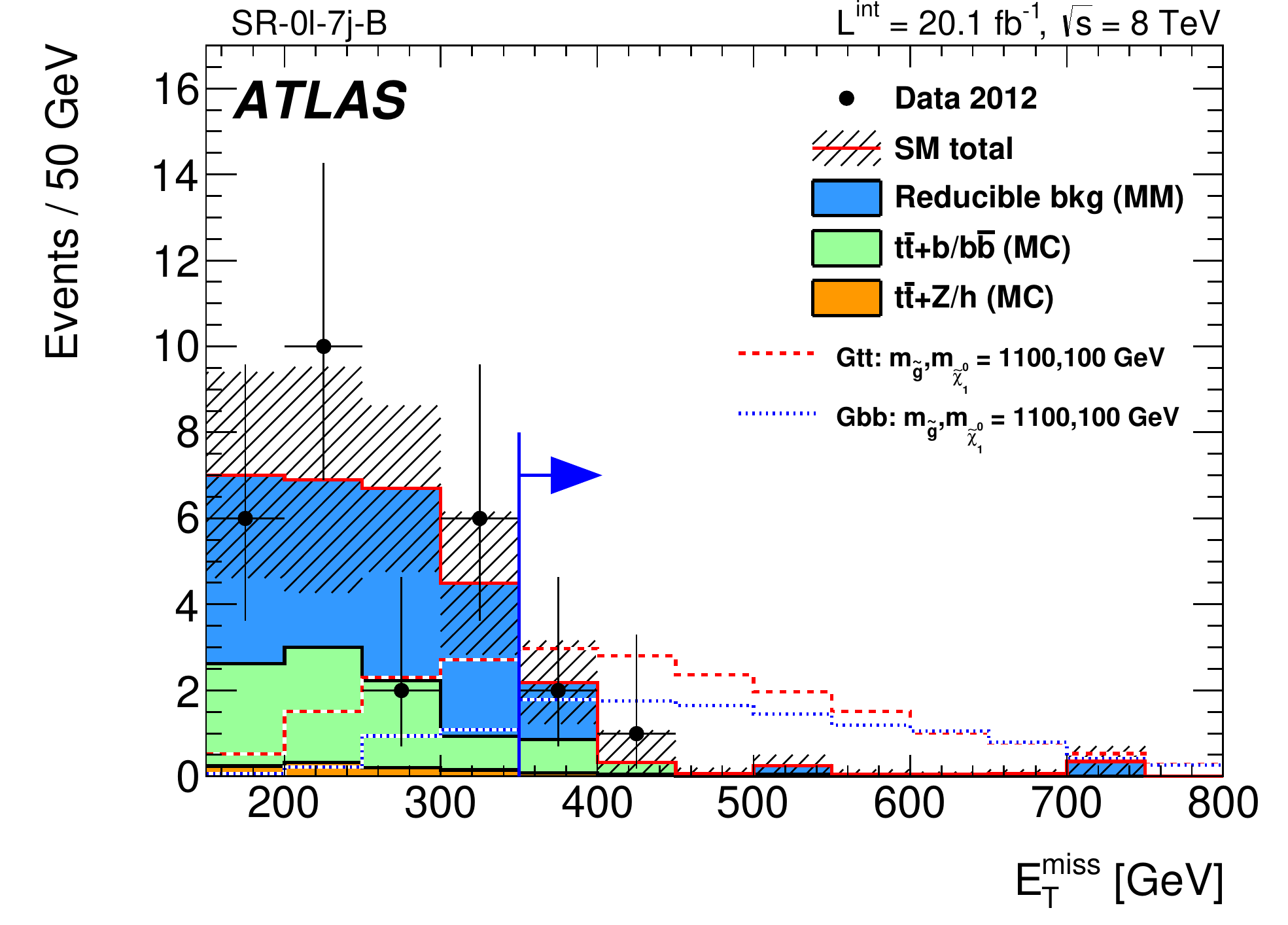}}  \\
\subfigure[]{\includegraphics[width=0.49\columnwidth]{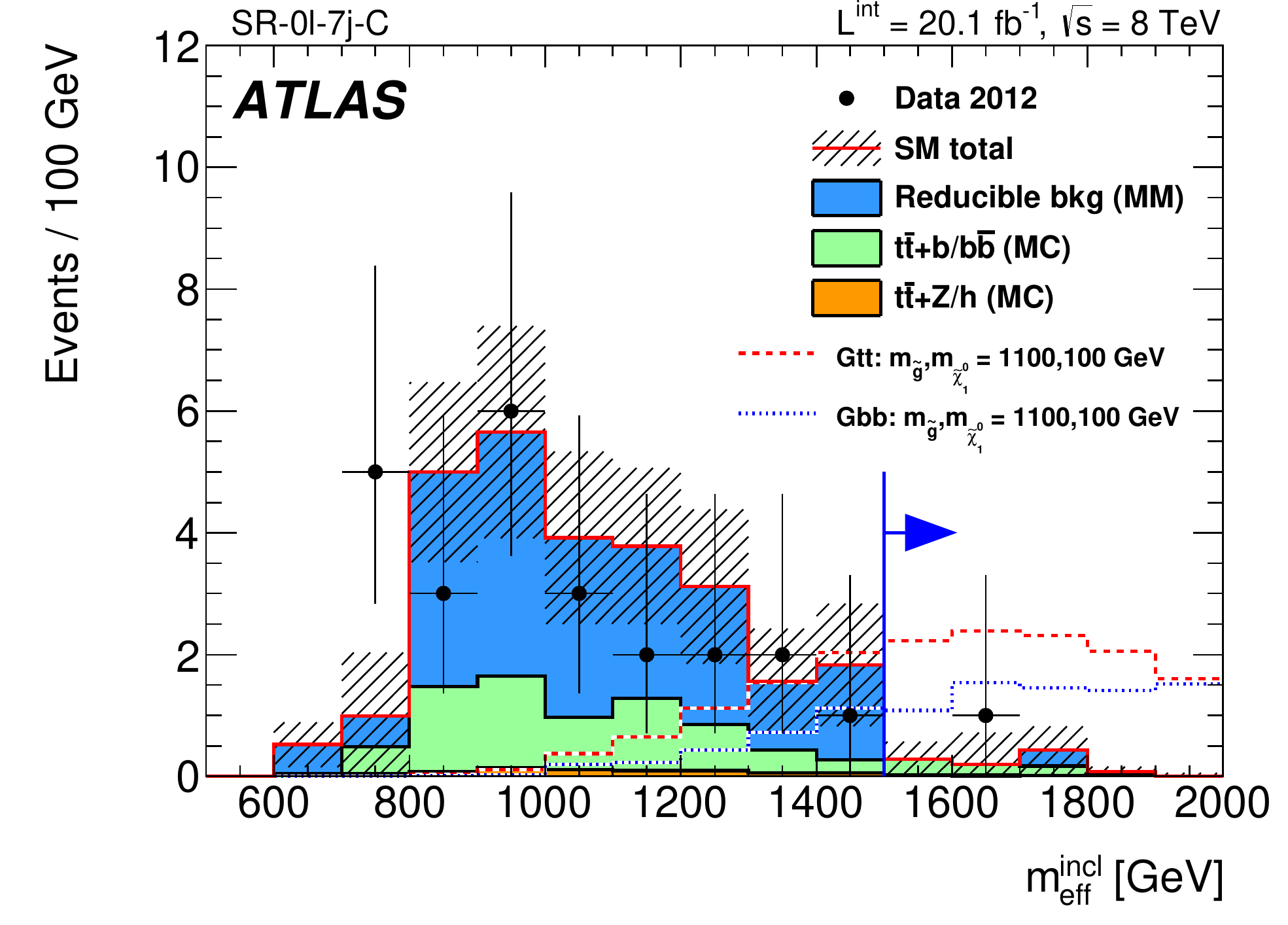}} \\ 
\caption{The (a)-(c) \meffi\   and (b) \met\  distributions observed in data for SR-0$\ell$-7j-A, SR-0$\ell$-7j-B and SR-0$\ell$-7j-C, respectively,  
after all requirements applied but the one indicated by the arrow, together with the background prediction. 
 The shaded bands include all  experimental systematic uncertainties on the background prediction.
 The prediction for two signal points from the Gtt \mbox{($\gl \rightarrow t\bar{t} \tilde{\chi}_1^0$)} and Gbb \mbox{($\gl \rightarrow b\bar{b} \tilde{\chi}_1^0$)}   models are overlaid.  
 The normalisation of the irreducible background \ttbb\ is as predicted by its theoretical
  cross-section scaled to the same luminosity as the data, prior to the fit in the control region. 
} 
\label{fig:SR_0l_7j}
\end{figure}
\clearpage

\begin{figure}[htp]
\centering
\subfigure[]{\includegraphics[width=0.49\columnwidth]{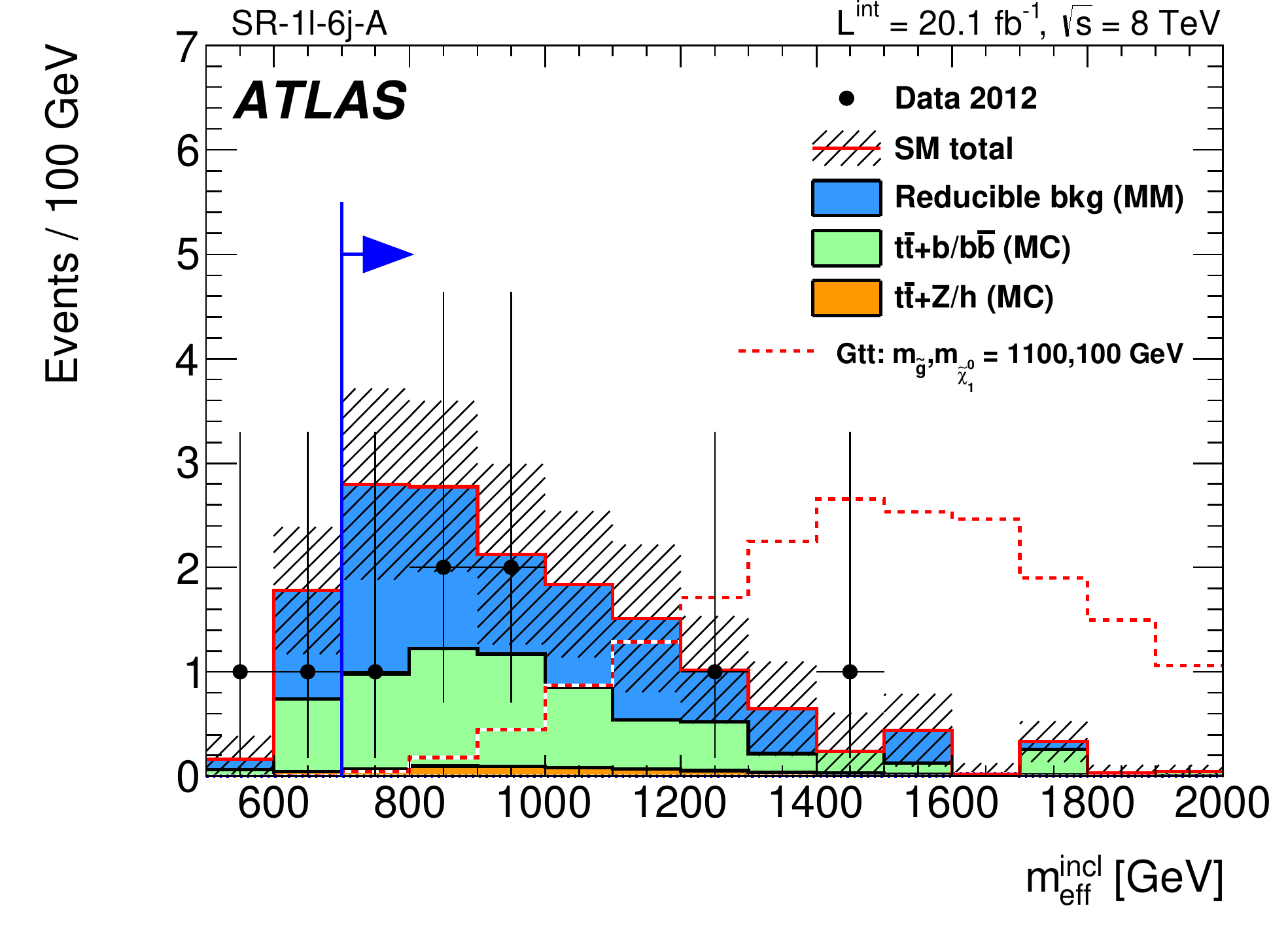}} 
\subfigure[]{\includegraphics[width=0.49\columnwidth]{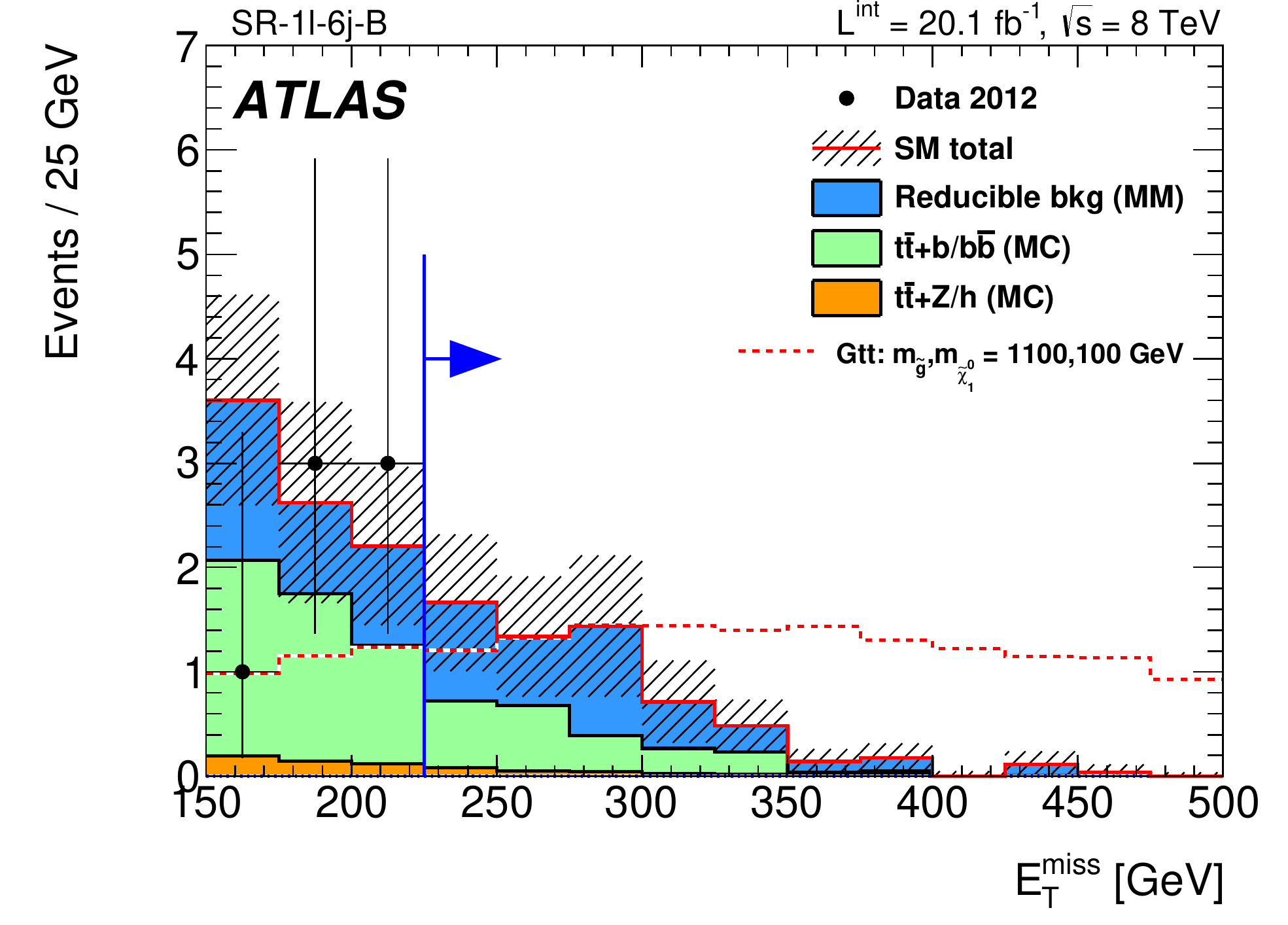}}  \\
\subfigure[]{\includegraphics[width=0.49\columnwidth]{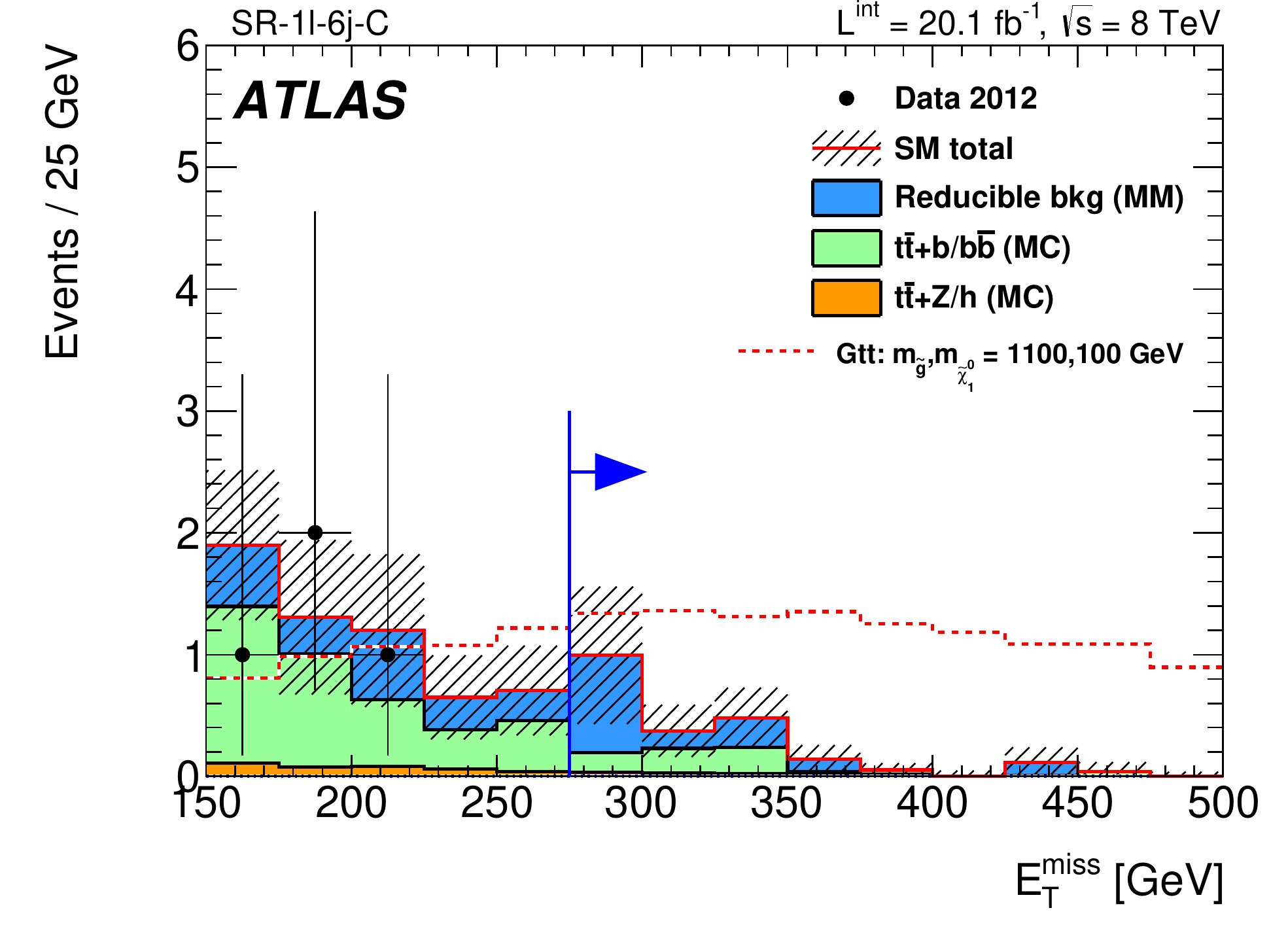}} \\ 
\caption{The (a) \meffi\  and (b)-(c) \met\  distribution observed in data for SR-1$\ell$-6j-A, SR-1$\ell$-6j-B and SR-1$\ell$-6j-C, respectively,  
after all requirements applied but the one indicated by the arrow, together with the background prediction. 
 The shaded bands include all  experimental systematic uncertainties on the background prediction.
The prediction for one signal point from the Gtt ($\gl \rightarrow t\bar{t} \tilde{\chi}_1^0$)  model is overlaid. 
 The normalisation of the irreducible background \ttbb\ is as predicted by its theoretical
  cross-section scaled to the same luminosity as the data, prior to the fit in the control region. 
} 
\label{fig:SR_1l}
\end{figure}

\clearpage

\begin{table*}[htp]
\centering
\begin{tabular}{lcccc}
\toprule
           & SR-0$\ell$-4j-A            &   SR-0$\ell$-4j-B          &   SR-0$\ell$-4j-C*        \\
\midrule
Observed events          &  $2$  &   $3$  &    $1$       \\
\midrule
Fitted background events              &   $1.6 \pm 0.9$ &   $1.3 \pm 0.9$   &    $1.6 \pm 0.7$    \\
\midrule
Reducible background  events       &     $1.1 \pm 0.8$   & $0.7_{-0.7}^{+0.8}$    &  $1.1 \pm 0.5$   \\
\ttbb\  events               &  $0.4 \pm 0.3$   & $0.4 \pm 0.3$  &  $0.4 \pm 0.4$ \\
\ttbar+($Z \rightarrow b\bar{b}$) events               &   $0.03 \pm 0.02$   & $0.04 \pm 0.03$    & $0.01 \pm 0.01$  \\          
\ttbar+($h \rightarrow b\bar{b}$) events         &   $0.05 \pm 0.05$   & $0.07 \pm 0.07$    &  $0.03 \pm  0.03$ \\          
\midrule
MC-only prediction & $2.6$ & $3.7$ & $2.3$ \\  
\midrule
$CL_b$ & $0.65$ & $0.87$ & $0.39$ \\ 
$p_0$ & $0.40$ & $0.15$ & $0.5$ \\ 
Expected UL on $N_{\mathrm{BSM}}$  & ${4.5}^{+1.7}_{-0.5}$ (${4.3}^{+2.6}_{-0.9}$) &  ${4.5}^{+1.7}_{-0.6}$ (${4.3}^{+2.6}_{-1.0}$) & ${4.1}^{+1.7}_{-0.3}$ (${4.1}^{+2.2}_{-1.0}$)    \\
Observed UL on $N_{\mathrm{BSM}}$  &  $5.2$ ($5.0$)  &   $6.5$ ($6.2$) &   $3.9$ (3.8) \\
Observed (expected) UL on $\sigma_{\mathrm{vis}}$ [fb] & 0.26 (0.22)   & 0.32 (0.22)   & 0.19 (0.20)    \\
\midrule
                & SR-0$\ell$-7j-A            &   SR-0$\ell$-7j-B           &   SR-0$\ell$-7j-C         \\
\midrule
Observed events                   & $21$  & $3$    & $1$  \\
\midrule
Fitted background events              & $21.2 \pm 4.6$   & $3.2 \pm 1.6$ & $0.9_{-0.9}^{+1.0}$   \\
\midrule
 Reducible background events        & $ 13.6 \pm 4.0$  & $1.7 \pm 1.2$ & $<0.65$   \\
\ttbb\ events            & $ 6.7 \pm 3.9$  & $1.3 \pm 1.1$  & $0.8 \pm 0.7$   \\
\ttbar+($Z \rightarrow b\bar{b}$) events    &   $ 0.3 \pm 0.1$    & $0.07 \pm 0.03$  &   $0.03 \pm 0.03$  \\          
\ttbar+($h \rightarrow b\bar{b}$) events   &   $ 0.5 \pm 0.5$   & $0.1 \pm 0.1$  &   $0.06 \pm 0.06$ \\         
 \midrule
MC-only prediction & $31.4$ & $6.8$ & $3.1$ \\  
\midrule 
$CL_b$ & $0.51$ & $0.53$ & $0.59$ \\ 
$p_0$ & $0.50$ & $0.50$ & $0.46$ \\ 
Expected UL on $N_{\mathrm{BSM}}$  & ${13.8}^{+4.7}_{-3.4}$ (${13.5}^{+5.3}_{-3.7}$) & ${6.0}^{+2.2}_{-1.3}$ (${5.9}^{+2.5}_{-1.8}$) & ${4.1}^{+1.6}_{-0.8}$ (${4.1}^{+2.1}_{-1.0}$) \\
Observed UL on $N_{\mathrm{BSM}}$  &  $13.9$ ($13.4$) & $6.1$ ($5.8$) &  $4.2$ ($4.1$) \\
Observed (expected) UL on $\sigma_{\mathrm{vis}}$ [fb] & 0.69 (0.69)   & 0.30 (0.30)   & 0.21 (0.20)    \\
\bottomrule
\end{tabular}
\caption{Results of the likelihood fit in all 0-lepton signal regions. The errors shown include all systematic uncertainties. 
The data in the signal regions are not included in the fit. 
The MC-only predictions are given for comparison.  
The $CL_b$-values, which  quantify the observed level of agreement with the expected yield, 
and the $p_0$-values, which represent the probability of the SM background alone to fluctuate to the observed number of events or higher, are also reported.
The $p_0$-values are truncated at 0.5 if the number of observed events is below the number of expected events.     
Also shown are the expected and observed upper limits (UL) at 95\% CL on the number of beyond-the-SM events $N_{\mathrm{BSM}}$   in each SR. 
These limits are derived with pseudo-experiments, and the results obtained  with an asymptotic approximation are given in parentheses for comparison. 
They are used to derive upper limits on the visible cross-section  $\sigma_{\mathrm{vis}} = \sigma\times A \times\epsilon$ for hypothetical non-SM contributions.
  }
   \label{t-SR0L}
\end{table*}

\begin{table*}[htp]
\centering
\begin{tabular}{lccc}
\toprule
        & SR-1$\ell$-6j-A            &   SR-1$\ell$-6j-B           &   SR-1$\ell$-6j-C         \\[-0.05cm]
\midrule
Observed events              & $7$    & $0$    & $0$   \\
\midrule
Fitted background events           &   $13.5 \pm 3.2$ & $6.1 \pm 1.8$  & $2.3 \pm 0.7$  \\
\midrule
 Reducible background events      &    $7.2 \pm 3.4$  & $3.7 \pm 1.9$   & $1.5 \pm 0.7$ \\
\ttbb\  events           & $5.7 \pm 3.1$ & $2.1 \pm 1.4$  & $0.7 \pm 0.5$ \\
 \ttbar+($Z \rightarrow b\bar{b}$) events   &   $0.2 \pm 0.1$    & $0.11 \pm 0.08$  &   $0.07 \pm 0.03$  \\          
 \ttbar+($h \rightarrow b\bar{b}$) events   &   $0.4 \pm 0.4$   & $0.2 \pm 0.2$  &   $0.08 \pm  0.08$ \\            
 \midrule
 MC-only prediction & $16.4$ & $7.5$ & $2.8$ \\  
\midrule
$CL_b$ & $0.10$ & $0.02$ & $0.17$ \\ 
$p_0$ & $0.50$ & $0.50$ & $0.50$ \\ 
Expected UL on $N_{\mathrm{BSM}}$  & ${9.0}^{+3.5}_{-2.3}$  (${9.1}^{+4.0}_{-2.8}$) & ${6.0}^{+2.3}_{-1.7}$ (${6.0}^{+3.0}_{-1.9}$) & ${4.3}^{+1.8}_{-0.5}$ (${4.2}^{+2.6}_{-0.9}$)  \\
Observed UL on $N_{\mathrm{BSM}}$  & $6.1$ ($5.8$) &  $3.5$ (3.2) &  $3.6$ (2.9)  \\
Observed (expected) UL on $\sigma_{\mathrm{vis}}$ [fb] & 0.30 (0.45)   & 0.17 (0.30)   & 0.18 (0.21)    \\
\bottomrule
\end{tabular}

\caption{Results of the likelihood fit in all 1-lepton signal regions. The errors shown include all systematic uncertainties. 
The data in the signal regions are not included in the fit. 
The MC-only predictions are given for comparison.  
The $CL_b$-values, which  quantify the observed level of agreement with the expected yield, 
and the $p_0$-values, which represent the probability of the SM background alone to fluctuate to the observed number of events or higher, are also reported.
The $p_0$-values are truncated at 0.5 if the number of observed events is below the number of expected events.     
Also shown are the expected and observed upper limits (UL) at 95\% CL on the number of beyond-the-SM events $N_{\mathrm{BSM}}$   in each SR. 
These limits are derived with pseudo-experiments and the results obtained  with an asymptotic approximation are given in parentheses for comparison. They are used to derive upper limits on the visible cross-section  $\sigma_{\mathrm{vis}} = \sigma\times A \times\epsilon$ for hypothetical non-SM contributions.
  }
\label{t-SR1L}
\end{table*}

\section{Interpretations} \label{sec-int}

The results are used to derive exclusion limits in the context of several SUSY models (see section~\ref{sec-susy})
including bottom quarks or top quarks in the decay chain.  
The expected and observed exclusion limits are calculated using the asymptotic  approximation for each  SUSY model, 
treating the systematic uncertainties  as fully correlated between the signal and the background and between the 0- and 1-lepton channels 
where appropriate, and including  the expected signal contamination in the CR.  
Theoretical uncertainties on the SUSY signals are estimated as described in section~\ref{sec-samples}.
Limits are calculated for the nominal cross-section, and for the 
$\pm 1 \sigma^{\mathrm{SUSY}}_{\mathrm{theory}}$ cross-sections. 
All limits quoted in the text correspond to the $-1 \sigma^{\mathrm{SUSY}}_{\mathrm{theory}}$ 
hypothesis.

Limits are derived using the SR yielding the best expected sensitivity for each point 
in the parameter space, derived prior to having considered the data in the SR. For signal models 
where both the 0- and 1-lepton channels contribute to the sensitivity, 
these are combined in a simultaneous fit to enhance the sensitivity 
of the analysis. In this case, all possible permutations between the three 1-lepton and the six 0-lepton SRs 
are considered for each point of the parameter space, and the best expected combination is used. 
The SR-0$\ell$-4j signal regions are mostly sensitive to the gluino decays 
$\gl \rightarrow b\bar{b}\neut$ via on-shell 
or off-shell sbottoms, whilst the SR-0$\ell$-7j and SR-1$\ell$-6j signal regions are used to set exclusion 
limits in models where top quark enriched final states are expected. 

The expected and observed 95\% CL exclusion limits obtained with the 0-lepton
 channel for the direct--sbottom model are presented in the ($m^{}_{\bone},m^{}_{\neuttwo}$)
 plane in figure~\ref{f-sbottom}. Sbottom masses between 340~\gev\ and 600~\gev\ are excluded for 
$m_{\neuttwo} = 300$~\gev. No sensitivity is obtained for 
low $m_{\neuttwo}$ due to the soft \met\ expected for these signal events.  
The sensitivity of this analysis to $\tilde{b}_1$ pair production processes where 
$\bone \rightarrow b+\neuttwo$, $\neuttwo \rightarrow h + \neut$, 
depends on $m_{\neut}$. For higher neutralino masses, the sensitivity decreases because 
of the tight \met\ and jet $\pt$ selections applied in this analysis.

\begin{figure}[h]
\centering
\includegraphics[width=0.49\columnwidth]{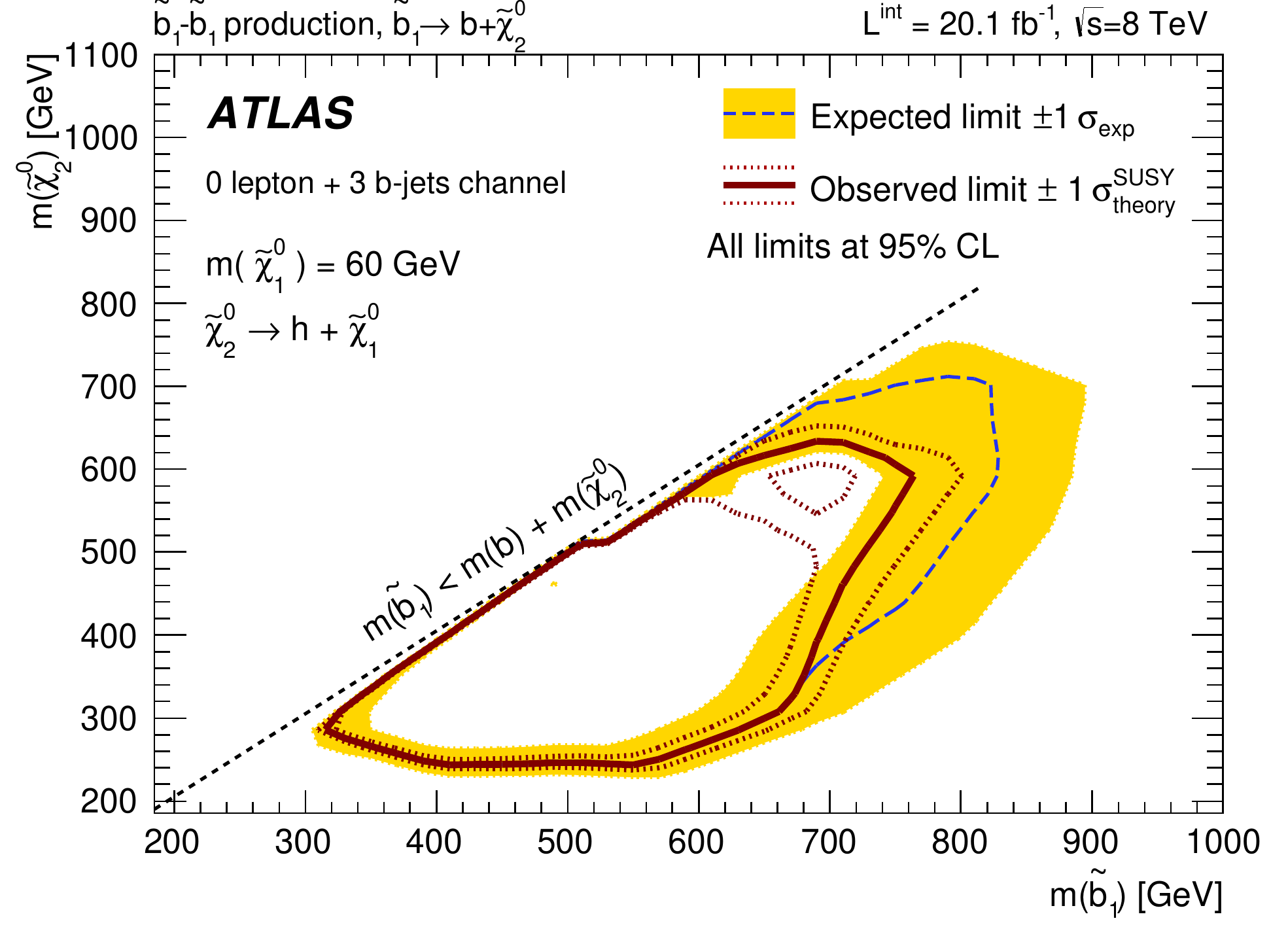} 
\caption{Exclusion limits in the  ($m_{\tilde{b}_1},m_{\tilde{\chi}_2^0}$) plane for the direct--sbottom model.  
The dashed blue and solid bold red lines show the 95\% CL expected and observed limits respectively, including all
uncertainties except the theoretical signal cross-section uncertainty. The shaded (yellow) bands around the expected limits show 
the impact of the experimental and background theoretical uncertainties while the dotted red 
lines show the impact on the observed limit of the variation of the nominal signal cross-section by 1$\sigma$ of its theoretical uncertainty. 
}
\label{f-sbottom}
\end{figure}

The expected and observed exclusion limits for the gluino--sbottom scenario are shown in 
figure~\ref{f-MSSM} (a). Exclusion limits are presented in the ($m^{}_{\gl},m^{}_{\bone}$) plane for the 0-lepton channel.
Gluino masses below 1250~\GeV\ are excluded for sbottom masses up to about 900~\GeV.
The result is complementary to the ATLAS search for direct sbottom pair production also carried out with 20.1 fb$^{-1}$ of data at 8~\tev~\cite{Aad:2013ija}. 

A combination of the 0- and 1-lepton results is used to derive the limit contours for the gluino--stop I and II models, 
presented in the ($m^{}_{\gl},m^{}_{\tone}$) plane in
figure~\ref{f-MSSM} (b) and (c). 
Gluino masses below 1180~\GeV\ are excluded for stop masses up to 1000~\GeV\ in the gluino--stop I model, 
while gluino masses below 1190~\GeV\ are excluded for stop masses up to 1000~\GeV\ in the gluino--stop II model. 
The sensitivity is lower in the  gluino--stop I model for most of the parameter space where soft \met\ and jets are expected from the chargino 
decay $\chipm\rightarrow W^*\neut$. 
The result is complementary to the ATLAS searches for direct stop pair production  performed in the 0-lepton channel with 20.1 fb$^{-1}$ of data at 8~\tev~\cite{Aad:2014bva} 
and in the  1-lepton channel with 4.7 fb$^{-1}$ of data at 7~\tev~\cite{Aad:2012xqa}.

\begin{figure*}[h]
\centering
\subfigure[]{\includegraphics[width=0.49\columnwidth]{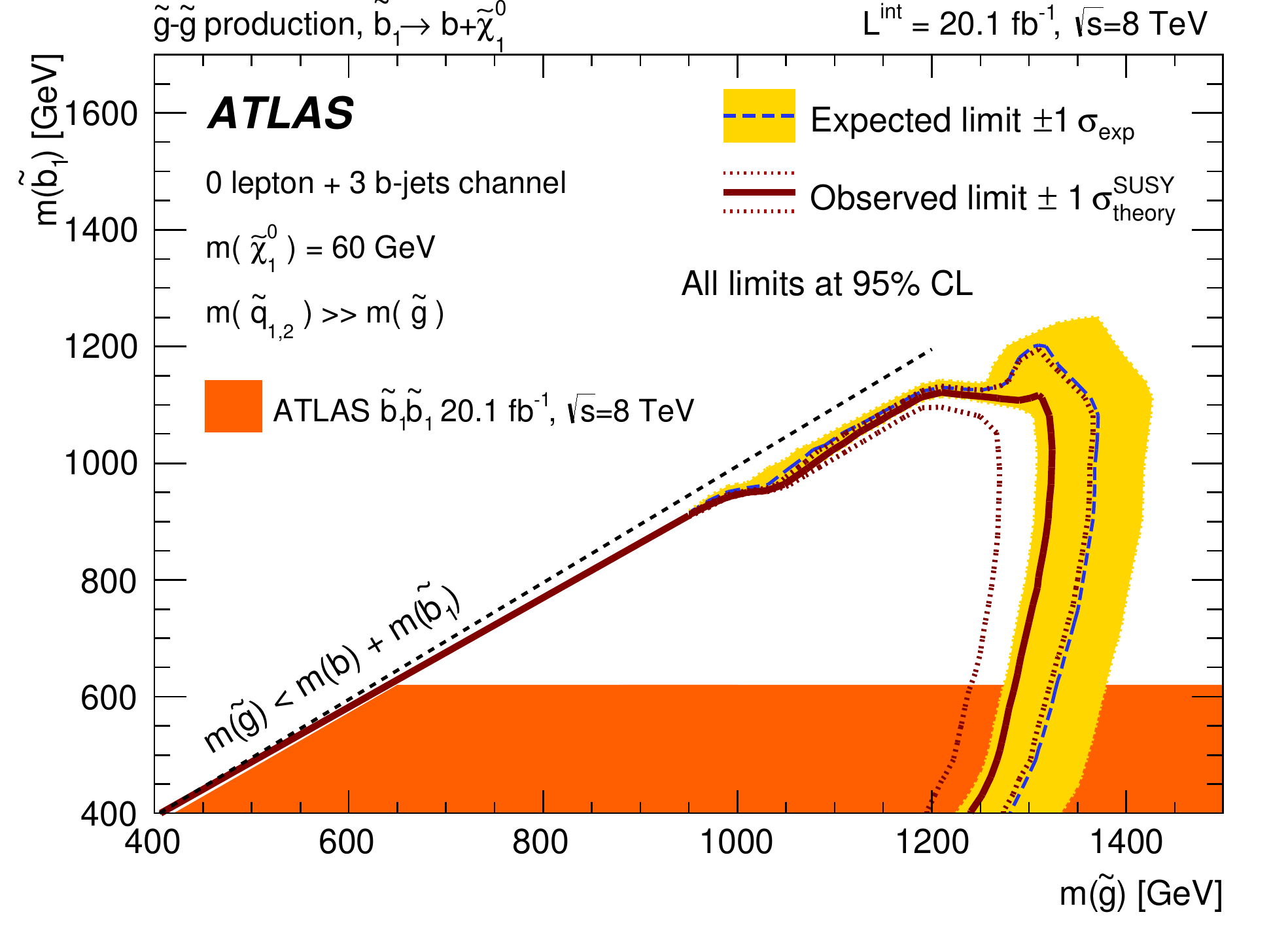}} \\
\subfigure[]{\includegraphics[width=0.49\columnwidth]{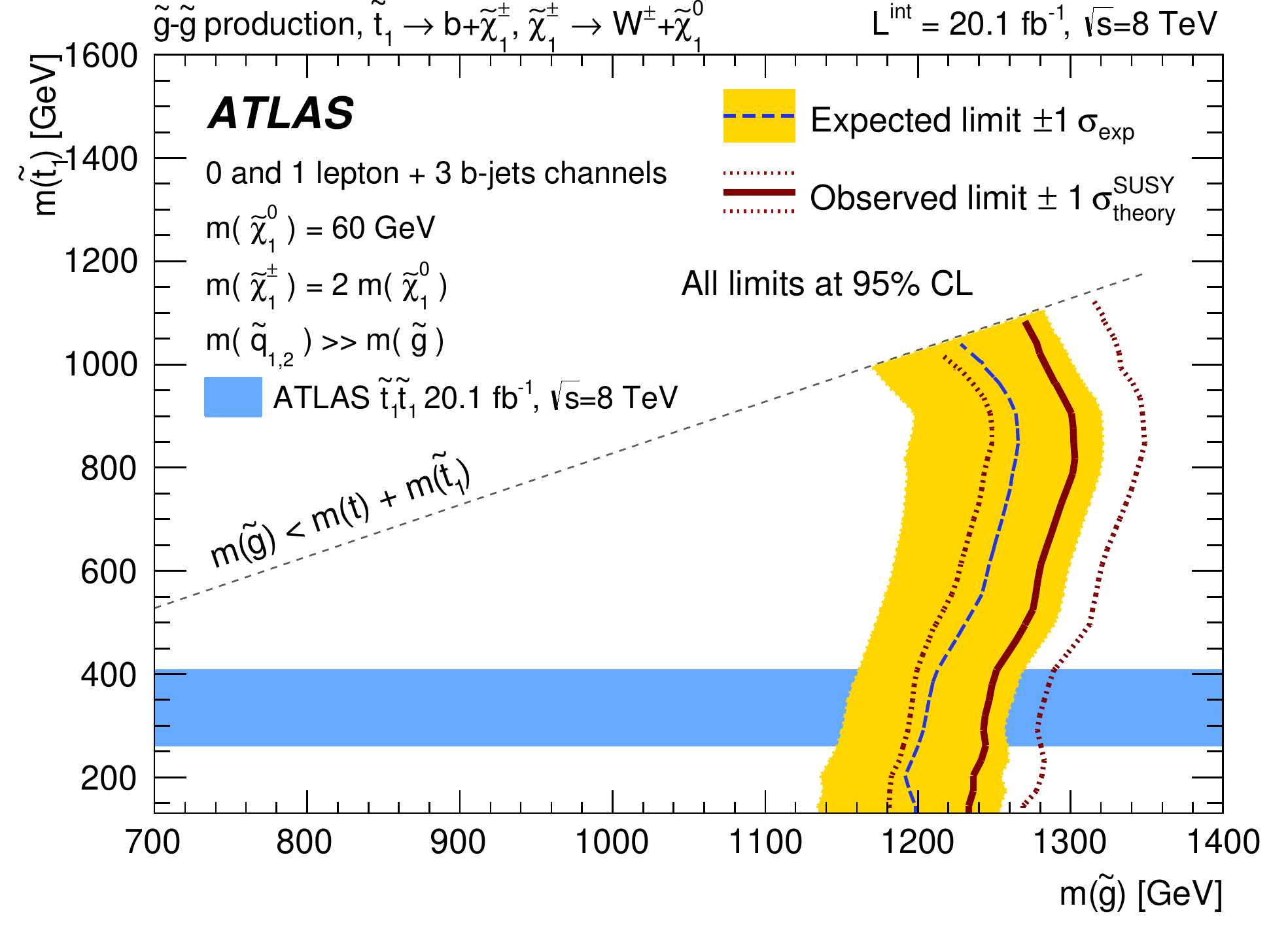}} 
\subfigure[]{\includegraphics[width=0.49\columnwidth]{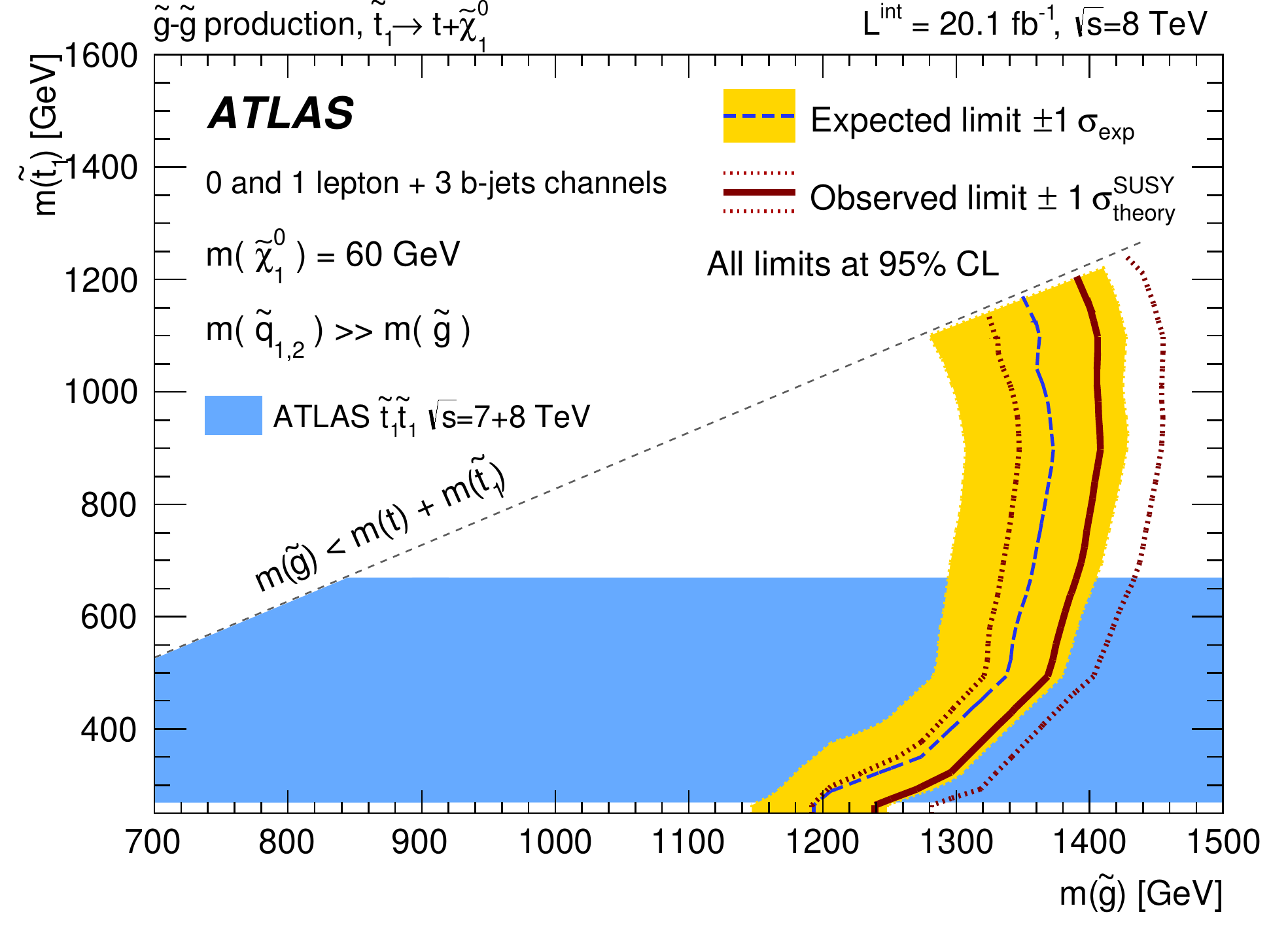}} 

\caption{Exclusion limits in the ($m_{\tilde{g}},m_{\tilde{b}_1}$) plane for the (a) gluino--sbottom model, and  
 in the  ($m_{\tilde{g}},m_{\tilde{t}_1}$) plane for the gluino--stop (b) I  and  (c) II models. 
The dashed blue and solid bold red lines show the 95\% CL expected and observed limits respectively, including all
uncertainties except the theoretical signal cross-section uncertainty. The shaded (yellow) bands around the expected limits show 
the impact of the experimental  and background theoretical uncertainties while the dotted red 
lines show the impact on the observed limit of the variation of the nominal signal cross-section by 1$\sigma$ of its theoretical uncertainty.
Also shown for reference are the results from the ATLAS sbottom and stop searches~\protect\cite{Aad:2013ija,Aad:2012xqa,Aad:2014bva} 
 derived using the nominal cross section. }
\label{f-MSSM}
\end{figure*}

The expected and observed exclusion limits for the Gbb model are shown in figure~\ref{fig:Gtt} (a). 
 As for the gluino--sbottom model,  four $b$-jets and \met\ are expected in the final state and only the 0-lepton channel is
used for the interpretation.  
Gluino masses below 1250~\GeV\ are excluded for $m_{\neut}<400$~\GeV\ while neutralino masses below 600~\GeV\ are excluded for 
  $m_{\tilde{g}}=1000$~\GeV. 
Lower sensitivity is achieved at very low mass splitting between the gluino and the neutralino 
because of the presence of soft $b$-jets and the low \met\ expected in signal events.

The combination of the 0-lepton and 1-lepton channels
is used to obtain the exclusion contours for the Gtt model, displayed in figure~\ref{fig:Gtt} (b). Gluino masses below 1340~\GeV\ 
are excluded for $m_{\neut}<400$~\GeV\ while neutralino masses below 620~\GeV\ are excluded for 
  $m_{\tilde{g}}=1000$~\GeV. The SR-0$\ell$-7j signal regions have the best sensitivity at large mass splitting 
between the gluino and the neutralino, where hard jets and large \met\ are expected, while the 1-lepton SRs 
have a better sensitivity close to the kinematic boundary.

Figure~\ref{fig:Gtt} (c) shows the expected and observed exclusion limits for
the Gtb scenario. The combination of the two channels is used to set the excluded area.
Gluino masses below 1300~\GeV\ are excluded for $m_{\neut}<300$~\GeV\ while neutralino 
masses  below 600~\GeV\ are excluded for $m_{\tilde{g}}=1100$~\GeV.

\begin{figure*}[h]
\centering
\subfigure[]{\includegraphics[width=0.49\columnwidth]{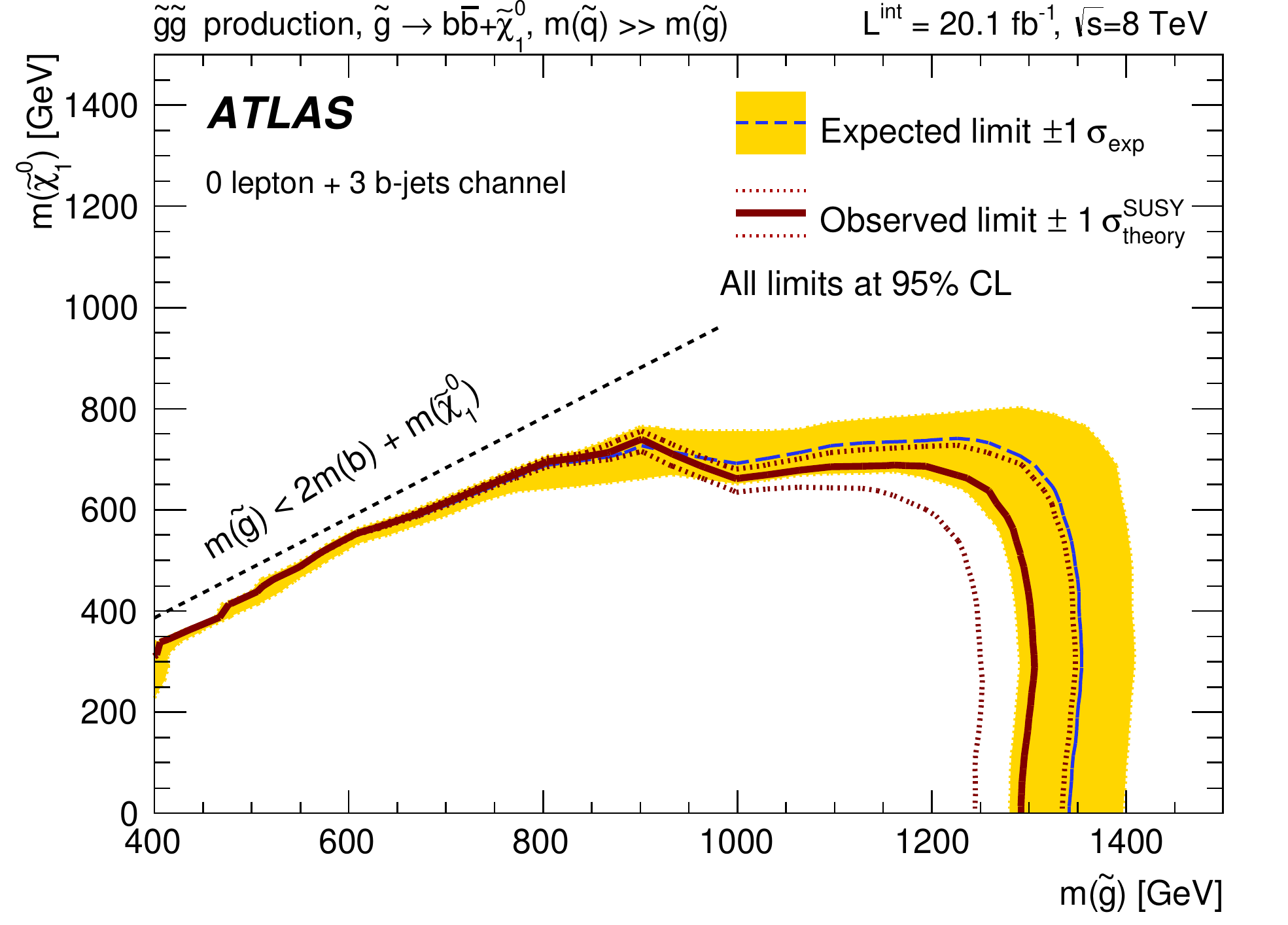}}
\subfigure[]{\includegraphics[width=0.49\columnwidth]{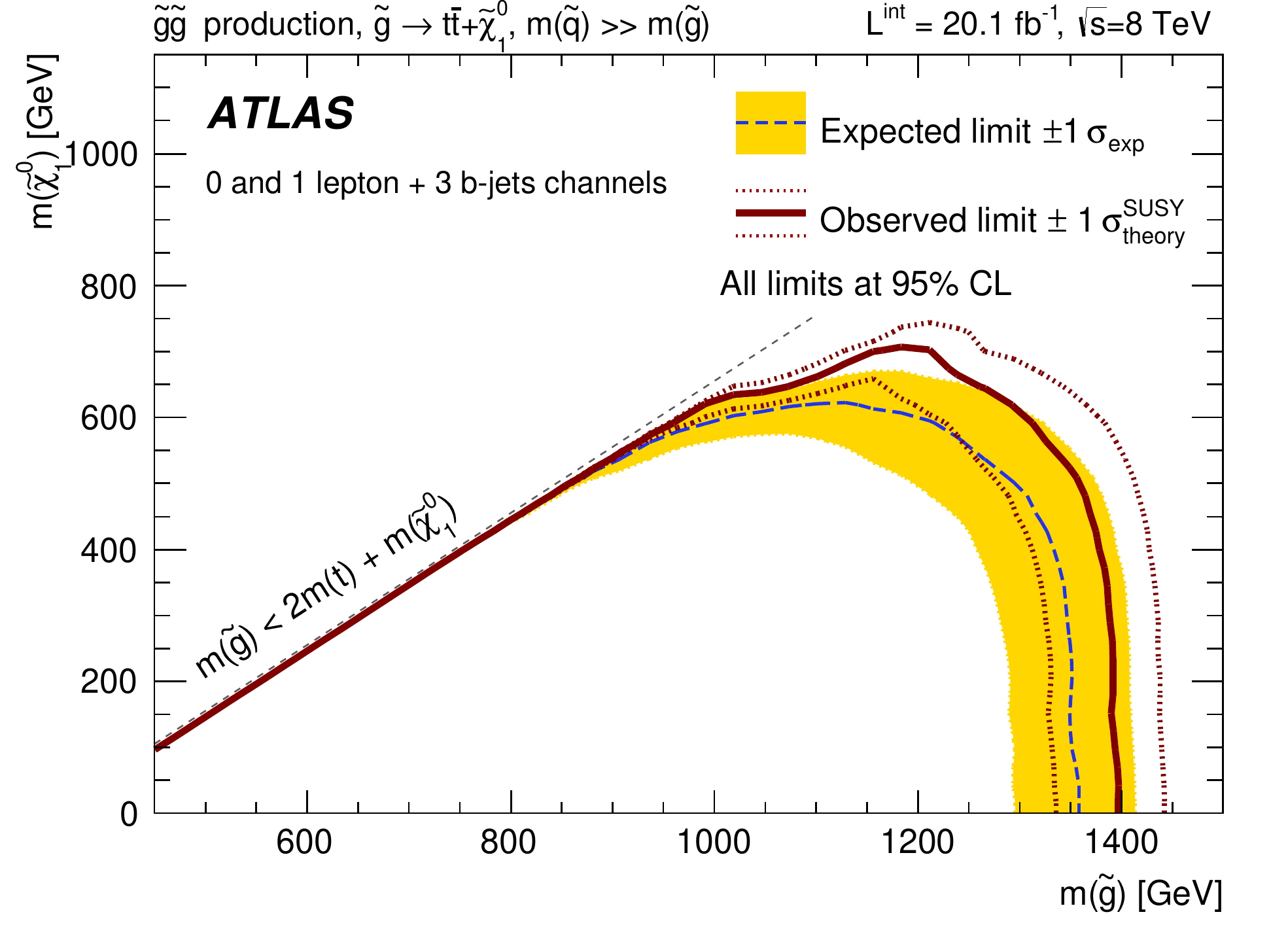}} 
\subfigure[]{\includegraphics[width=0.49\columnwidth]{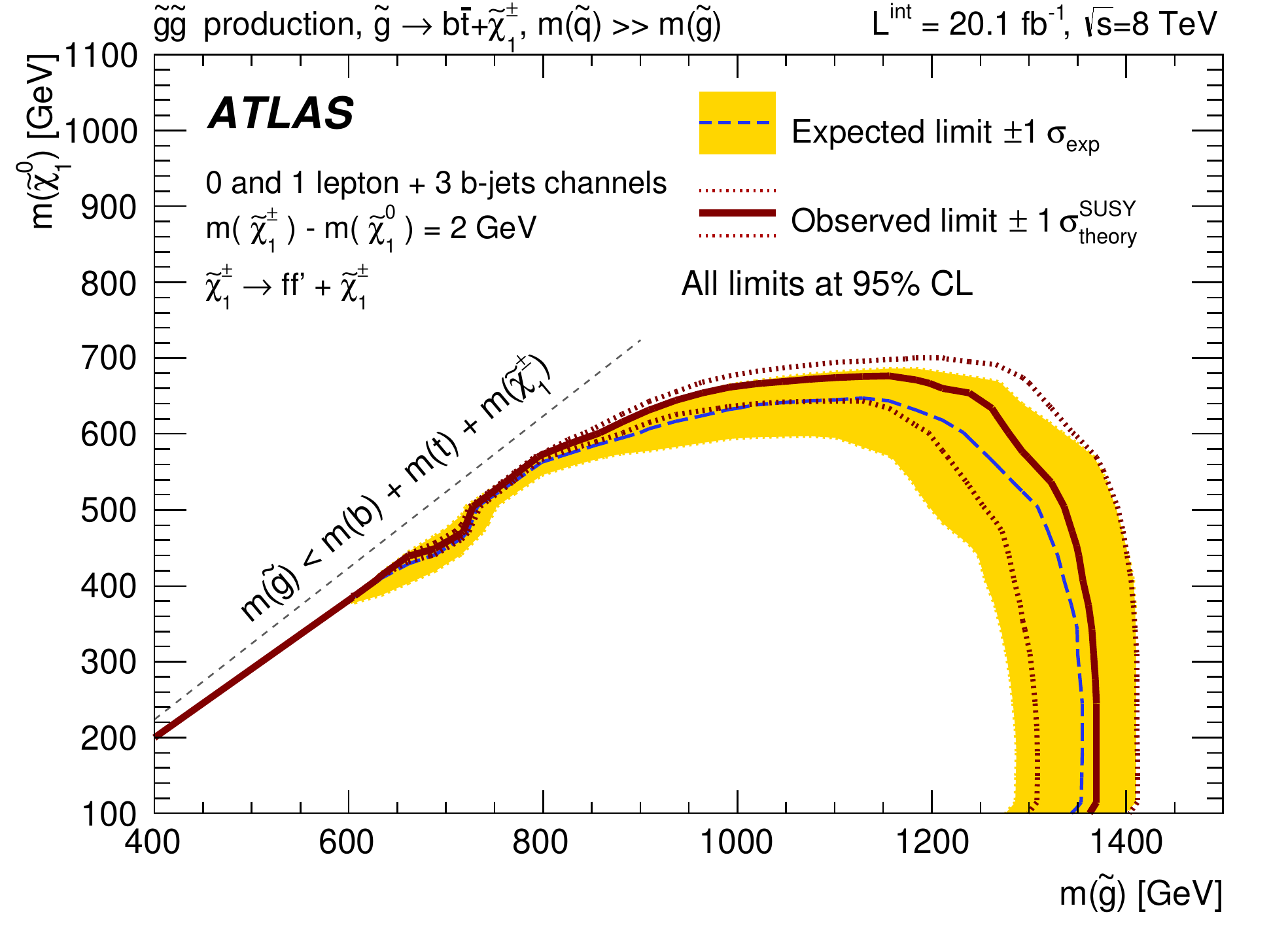}} 
\caption{
Exclusion limits in the  ($m_{\tilde{g}},m_{\tilde{\chi}_1^0}$) plane for the (a) Gbb, (b) Gtt and (c) Gtb  models. 
The dashed blue and solid bold red lines show the 95\% CL expected and observed limits respectively, including all
uncertainties except the theoretical signal cross-section uncertainty. The shaded (yellow) bands around the expected limits show 
the impact of the experimental  and background theoretical uncertainties while the dotted red 
lines show the impact on the observed limit of the variation of the nominal signal cross-section by 1$\sigma$ of its theoretical uncertainty.}
\label{fig:Gtt}
\end{figure*}

Finally, expected and observed 95\% CL limits for the mSUGRA/CMSSM scenario discussed in
section~\ref{sec-susy} are presented in the ($m_0$, $m_{1/2}$) 
plane in figure~\ref{fig:limits}. 
Gluino masses smaller than 1280~\GeV\ are excluded. 
This analysis is especially sensitive to the high $m_0$ region, where final states with four top quarks dominate.

\begin{figure}
\centering
\includegraphics[width=0.49\columnwidth]{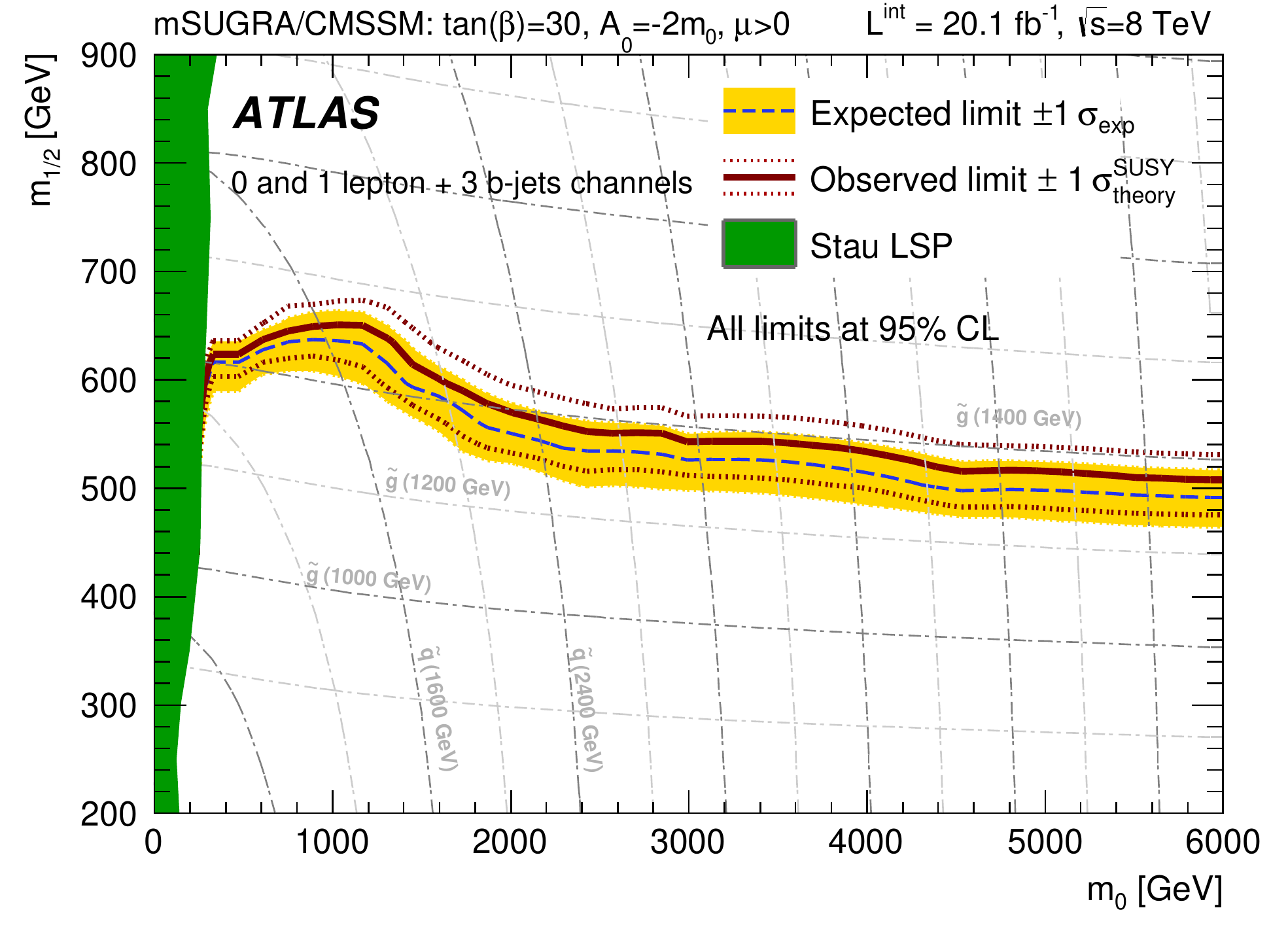} \\
\caption{
Exclusion limits in the  ($m_{0},m_{1/2}$) plane for the mSUGRA/CMSSM model. 
The dashed blue and solid bold red lines show the 95\% CL expected and observed limits respectively, including all
uncertainties except the theoretical signal cross-section uncertainty. The shaded (yellow) bands around the expected limits show 
the impact of the experimental  and background theoretical uncertainties while the dotted red 
lines show the impact on the observed limit of the variation of the nominal signal cross-section by 1$\sigma$ of its theoretical uncertainty.}
\label{fig:limits}
\end{figure}

\section{Conclusions}\label{sec-conclusion}

A search is presented in this paper for pair production of gluinos and sbottoms decaying into final states with multi-$b$-jets and 
missing transverse momentum.  This analysis uses \lumi~$\ifb$ of $pp$ collisions at a centre-of-mass energy of  8~\TeV\ collected by the ATLAS
experiment at the LHC. Events with large missing transverse momentum, at least four to at least seven jets, and at least three $b$-jets are considered.
The analysis is carried out separately for events with and without leptons in the final state, and the two channels are combined to enhance the sensitivity 
to SUSY scenarios with top quarks in the decay chain. 
No significant excess of events above SM expectations is found in data and the results are interpreted in the context of various simplified models 
involving gluinos, sbottoms and stops. In particular,  gluino masses up to about 1340~\GeV\ are excluded at 95\%~CL  in some models.

\FloatBarrier


\section*{Acknowledgments}




We thank CERN for the very successful operation of the LHC, as well as the
support staff from our institutions without whom ATLAS could not be
operated efficiently.

We acknowledge the support of ANPCyT, Argentina; YerPhI, Armenia; ARC,
Australia; BMWF and FWF, Austria; ANAS, Azerbaijan; SSTC, Belarus; CNPq and FAPESP,
Brazil; NSERC, NRC and CFI, Canada; CERN; CONICYT, Chile; CAS, MOST and NSFC,
China; COLCIENCIAS, Colombia; MSMT CR, MPO CR and VSC CR, Czech Republic;
DNRF, DNSRC and Lundbeck Foundation, Denmark; EPLANET, ERC and NSRF, European Union;
IN2P3-CNRS, CEA-DSM/IRFU, France; GNSF, Georgia; BMBF, DFG, HGF, MPG and AvH
Foundation, Germany; GSRT and NSRF, Greece; ISF, MINERVA, GIF, I-CORE and Benoziyo Center,
Israel; INFN, Italy; MEXT and JSPS, Japan; CNRST, Morocco; FOM and NWO,
Netherlands; BRF and RCN, Norway; MNiSW and NCN, Poland; GRICES and FCT, Portugal; MNE/IFA, Romania; MES of Russia and ROSATOM, Russian Federation; JINR; MSTD,
Serbia; MSSR, Slovakia; ARRS and MIZ\v{S}, Slovenia; DST/NRF, South Africa;
MINECO, Spain; SRC and Wallenberg Foundation, Sweden; SER, SNSF and Cantons of
Bern and Geneva, Switzerland; NSC, Taiwan; TAEK, Turkey; STFC, the Royal
Society and Leverhulme Trust, United Kingdom; DOE and NSF, United States of
America.

The crucial computing support from all WLCG partners is acknowledged
gratefully, in particular from CERN and the ATLAS Tier-1 facilities at
TRIUMF (Canada), NDGF (Denmark, Norway, Sweden), CC-IN2P3 (France),
KIT/GridKA (Germany), INFN-CNAF (Italy), NL-T1 (Netherlands), PIC (Spain),
ASGC (Taiwan), RAL (UK) and BNL (USA) and in the Tier-2 facilities
worldwide.

\bibliographystyle{JHEP}
\raggedright 
\bibliography{paper}

\clearpage
\begin{flushleft}
{\Large The ATLAS Collaboration}

\bigskip

G.~Aad$^{\rm 84}$,
B.~Abbott$^{\rm 112}$,
J.~Abdallah$^{\rm 152}$,
S.~Abdel~Khalek$^{\rm 116}$,
O.~Abdinov$^{\rm 11}$,
R.~Aben$^{\rm 106}$,
B.~Abi$^{\rm 113}$,
M.~Abolins$^{\rm 89}$,
O.S.~AbouZeid$^{\rm 159}$,
H.~Abramowicz$^{\rm 154}$
% ...
\bigskip
\\
% ...
$^{11}$ Institute of Physics, Azerbaijan Academy of Sciences, Baku, Azerbaijan\\
% ...
$^{84}$ CPPM, Aix-Marseille Universit{\'e} and CNRS/IN2P3, Marseille, France\\
% ...
$^{89}$ Department of Physics and Astronomy, Michigan State University, East Lansing MI, United States of America\\
% ...
$^{106}$ Nikhef National Institute for Subatomic Physics and University of Amsterdam, Amsterdam, Netherlands\\
% ...
$^{112}$ Homer L. Dodge Department of Physics and Astronomy, University of Oklahoma, Norman OK, United States of America\\
$^{113}$ Department of Physics, Oklahoma State University, Stillwater OK, United States of America\\
% ...
$^{116}$ LAL, Universit{\'e} Paris-Sud and CNRS/IN2P3, Orsay, France\\
% ...
$^{152}$ Institute of Physics, Academia Sinica, Taipei, Taiwan\\
% ...
$^{154}$ Raymond and Beverly Sackler School of Physics and Astronomy, Tel Aviv University, Tel Aviv, Israel\\
% ...
$^{159}$ Department of Physics, University of Toronto, Toronto ON, Canada\\
% ...
\end{flushleft}

\end{document}